\documentclass[amssymb,amsmath,superscriptaddress,a4paper]{revtex4-1}

\usepackage{bbold}

\begin{document}

\title[From charge motion to the gyrokinetic equation]{ From charge motion in general magnetic fields\\ to the non perturbative gyrokinetic equation}

\author{C. Di Troia}

\affiliation{ENEA Unit\`{a} tecnica Fusione, C.R. Frascati, Via E. Fermi 45, 00044 Frascati (Rome), Italy}
\email{ claudio.ditroia@enea.it}
\vspace{10pt}

\begin{abstract}
The exact analytical description of non relativistic charge motion in general magnetic fields is, apparently, a simple problem but, it has not  been solved up to now apart for rare cases. The key feature of the present derivation is to adopt a non perturbative magnetic field description to find new solutions of motion. Among all solutions, two are particularly important:  guiding particle and gyro-particle solutions. The guiding particle has been characterized to be minimally coupled to the magnetic field, the gyro-particle has been defined to be maximally coupled to the magnetic field and, also, to move on a closed orbit. The generic charged particle motion is shown to be expressed as the sum of such particular solutions. This non perturbative approach corresponds to the description of the particle motion in the gyro-center  and/or guiding center reference frame obtained at all the orders of  the modern gyro-center transformation. The \emph{Boltzmann} equation is analyzed with the described exact guiding center coordinates. The obtained gyrokinetic equation is solved for the \emph{Boltzmann} equation at \emph{marginal stability} conditions.   
\end{abstract}
\maketitle

\\
\vspace{2pc}

\section{Introduction}
\label{sec:intro}

 Gyrokinetics is a theoretical  framework used for describing either astrophysical or laboratory plasmas behavior in the presence of magnetic fields. Its foundation is relied to the motion of a charged particle in a magnetic field. The trajectory of a charge is known to be a sort of helix moving circularly around a curve. A point on such curve is considered the origin of a spatial reference frame where such gyrating motion is fine described. Commonly the treatment is intrinsically perturbative: the exact solution is known for a uniform and constant magnetic field and such solution is thought as the 0$^{th}$ order solution. In a seminal paper \cite{littlejohn0} of the modern guiding center transformation, the first equation, here reported for convenience: 
\begin{equation}
\label{lj1}
\frac{d x^i}{d t}= X^i({\bf x}) =\sum_{k=0}^\infty \epsilon^k X^i_k({\bf x}),
\end{equation}
indicates how  intrinsic is the perturbative expansion in gyrokinetic theory. The left equivalence is the fundamental equation for a dynamical system (without time dependency) and it will be the starting equation also here (with the velocity vector field  indicated with $V$ instead of $X$). The right equivalence in (\ref{lj1}) is always considered in modern gyrokinetic theory, expressing a sort of intrinsic expansion useful
 when the magnetic field is slightly curved and stretched and, if time is considered,  also time fluctuations are present.  The consequence is that particles orbits are described with a certain degree of accuracy. Such procedure has been formalized with a huge mathematical effort to be extended at the wanted order. The gyrokinetic transformation is such expansion \cite{brizardhahmRMP}. High order expansion respect to the leading one must be performed  when the magnetic field shows an high level of   inhomogeneities in space and fluctuations in time.  However, this expansion is rarely seen at an order greater than the first. 
In the present approach there is no need of the right equivalence in (\ref{lj1}) because the magnetic field, $B$, is not perturbatively approached. $B$ is directly considered  nonhomogeneus and fluctuating from the beginning. 

Commonly there is an intermediate transformation introducing gyrokinetics: the Guiding Center (GC) transformation, which is mainly used when the magnetic field is slowly varying and the time evolution can be neglected (as in (\ref{lj1})).
The reference case is, always, the charge motion in a constant and uniform magnetic field. This example clarifies many things and also what the GC is. In a cartesian reference frame the magnetic field is   $B=B_0 e_z$ and the solution of motion is a velocity decomposed in a parallel  and a perpendicular to $B$ component: $v=v_\| e_z + v_\perp$. The perpendicular velocity can be expressed as $v_\perp=-| v_\perp | ( e_x \sin \gamma +e_y \cos \gamma)$, where $\gamma=\omega_c t$ is the \emph{gyro-angle} and  $\omega_c$ is the \emph{cyclotron} frequency, $\omega_c=eB_0/m$, where $e$ is the charge, positive for ions, negative for electrons, and $m$ is the mass of the considered species (the speed of light is $c=1$). The former velocity, $v$, satisfies  the non relativistic \emph{Lorentz} force law:
$$
\dot{v}= \frac{e}{m} (E+v \times B),
$$ 
when, as for example, the electric field $E=E_0 e_z$ is constant and parallel to $B$. This is easily proved substituting $\dot{v}_\|=eE_0/m$. The motion of the particle is the helix resulting from the uniformly circular motion of the vector \emph{Larmor} radius $\rho$ around the GC position $X$ which proceeds parallel to $B$ with velocity $v_\|$.  The particle position is expressed by $x=X+\rho$, where $\rho=|\rho|  (e_x \cos \gamma - e_y \sin \gamma)$ has constant magnitude $|\rho|=\rho_L=\pm|v_\perp|/\omega_c$ (the sign depends on $\omega_c$: positive if $e>0$, negative if $e<0$).
It is worth noticing that, although it is unstable, the GC itself describes a solution of motion and it can represent a real charged particle which is here called Guiding Particle (GP). Such particular solution has GC position $X$ and GC velocity $V=v_\| e_z$. Note that the GP motion ignores the presence of $B$ because it moves parallel to it. If the electric field intensity depends on space, the solution is similar but now $V=V(x)$ becomes a velocity vector field. The following three properties will be generalized for several $B$ fields: 1) GC can represent a particle obeying the \emph{Lorentz} force law, 2) thus GP is minimally coupled to the magnetic field, and 3) $V$ is a velocity vector field defined at each position (and at all times).\\
In the uniform and constant $B$ case, the motion is described in the reference frame centered at the GC position and velocity, $X$ and $V$, respectively. In such reference frame the (unperturbed) trajectory makes a loop, a circular motion which is well described by the radial vector $\rho$ circularly moving with velocity $v_\perp=\dot{\rho}$. This is again the leading example for the \emph{gyro-particle} motion. The gyro-particle, contrary to the GP, is maximally coupled to $B$ and it moves in a closed loop. These two properties will be considered also for the general magnetic field case. \\

For the present studies, it is convenient to slightly modify the formal expression describing the particle velocity. In the constant and uniform magnetic field case the velocity is 
\begin{equation}
\label{notation}
v=v_\| e_z+ v_\perp =v_\| \nabla z +\omega_c \nabla S \times \nabla z,
\end{equation}
being $\nabla z =e_z$ the direction of the magnetic field and, for $S=\rho^2/2$, $\nabla S= \rho$, the gyro-radius.  The above representation can be written as $v=v_\|\nabla z+\omega_c \nabla \times S \nabla z$, and considered a sort of \emph{Helmoltz} representation of $v$. The first component refers to the GP, whilst the last to the gyro-particle,  orthogonal either to the GP motion and  to $\nabla S$. Note that $\nabla S \cdot \nabla z =0$ then $\dot S= v\cdot \nabla S=0$: the surface $S$ is a constant of motion. 
The representation $(\ref{notation})$ is useful on visualizing the helical motion because a helicoid is the sum of two parts: a guiding curve, which for the uniform case  is a straight line, traced by $v_\| \nabla z$, and another curve winding around it. 
If general $B$ is analyzed, the angular velocity vector $\omega_c \nabla z$ becomes $\omega_c \nabla z \to \Omega$, and the GP velocity $v_\|\nabla z \to V$, but the same representation allow to recognize the footprint of the helicoidal motion.

\section{Motion of a charged particle in a general  magnetic field}
\label{sec:sec2}

The most important equation for a dynamical, time dependent, system is
\begin{equation}
\label{first}
\dot{X}=V(t,X).
\end{equation}
If the velocity vector field $V$ is given, the \emph{integral curve} $X$ could be obtained. Suppose that the field $V$ is unknown and you are looking for a solution $V$ which satisfies the non relativistic $Lorentz$ force law:
\begin{equation}
\label{2}
\dot{V} =\frac{e}{m}(E+V \times B),
\end{equation}
being $B=B(t,X)$, the external magnetic field, already known in terms of the vector potential $A=A(t,X)$, being $B=\nabla \times A$. The electric field, $E=-\nabla \Phi -\partial_t A$, being $\Phi=\Phi(t,X)$ the electric potential, is leaved undetermined.

From the vectorial identity
\begin{equation}
\label{vectId0}
\nabla (\mathrm{a}\cdot \mathrm{b})= \mathrm{a} \cdot \nabla \mathrm{b}+\mathrm{b}\cdot \nabla \mathrm{a} +\mathrm{a} \times \nabla \times \mathrm{b} + \mathrm{b} \times \nabla \times \mathrm{a}
\end{equation}
applied for $\mathrm{a}=\mathrm{b}=V$,
the explicit expression of $\dot{V}$ is
\begin{equation}
\label{3}
\dot{V}(X,t)=\partial_t V+V\cdot \nabla V=\partial_t V+\nabla V^2/2 - V \times (\nabla \times V).
\end{equation}
Using the electric and vector potential description of fields in (\ref{2}) and comparing the result with (\ref{3}), it is obtained the relation:
\begin{equation}
\label{4}
\partial_t V+\nabla V^2/2 - V \times (\nabla \times V)=-\frac{e}{m}(\partial_t A + \nabla \Phi -V\times B).
\end{equation}
Defining the moment per unit mass $P \equiv V+eA/m$, and the energy per unit mass $\mathcal{E} \equiv V^2/2+e\Phi/m$, the above equation becomes
\begin{equation}
\label{good}
\partial_t P+\nabla \mathcal{E}= V \times \left(\nabla \times V+\frac{e}{m}B \right).
\end{equation}
It follows that
  \begin{equation}
\label{good2}
V \cdot \partial_t P+V\cdot \nabla \mathcal{E}= 0.
\end{equation}
The considered system is known to be \emph{hamiltonian} (the \emph{Lorentz}'s law is the general \emph{Hamilton}'s equation for a conservative system with a velocity dependent potential energy \cite{engels,krupkova}) so that it is reasonable setting $\dot{\mathcal{E}}=\partial_t \mathcal{E}$, or $V\cdot \nabla \mathcal{E}= 0$. From (\ref{good2}), $ V \cdot \partial_t P=0$, or
 $V\cdot \partial_t V=-(e/m) V \cdot \partial_t A$, so that  
\begin{equation}
\label{dote}
\dot{\mathcal{E}}=V \cdot \partial_t V +(e/m)\partial_t \Phi=(e/m) \left(\partial_t \Phi -V \cdot \partial_t A \right).
\end{equation}
Asking for \emph{gauge} symmetry, $P\to P^\prime=V+eA^\prime/m=V+eA/m+e\nabla g/m$ and $\mathcal{E} \to \mathcal{E}^\prime=V^2/2+e\Phi^\prime/m=V^2/2+e\Phi/m-e\partial_t g /m$, for a $gauge$ function $g$. The $gauge$ symmetry ensures the invariance of (\ref{good}), being
\begin{equation}
\partial_t P+\nabla \mathcal{E}=\partial_t P^\prime+\nabla \mathcal{E}^\prime.
\end{equation}

The equation (\ref{good}) is related to the \emph{Euler-Lagrange} equation. In the \emph{lagrangian} formulation, the velocity is an unknown variable whilst, in the present approach, it is an unknown function of position and time. There is a  correspondence between the two approaches due to the fact that the velocity field must satisfies  the \emph{Lorentz} force law which can also be obtained from the \emph{Euler-Lagrange} equation. Analogously, the equation (\ref{good}) is not a \emph{Hamilton}'s equation although the system is \emph{hamiltonian}. It is worth to note that if you substitute $eA/m\to P-V$ and $e\Phi/m\to\mathcal{E}-V^2/2$ in the \emph{Lagrangian}, $L(X,\dot X,t)$, for the description of a charged particle in an e.m. field,  then  $L=\dot X^2/2+(eA/m)\cdot \dot X -e\Phi/m\to(\dot X-V)^2/2+P\cdot \dot X -\mathcal{E}$. When the motion is described in the proper reference frame then the kinetic energy is zero because the particle is at rest; this case is obtained by posing $\dot X =V(t,X)$, which is the equation (\ref{first}), and the \emph{Lagrangian} becomes $L=P\cdot V-\mathcal{E}$, coherently with the \emph{Legendre} transformation if the energy $\mathcal{E}$ is the  \emph{Hamiltonian} and $P$ is the momentum. 

The equation (\ref{good}) is transformed into the following system of equations:
\begin{equation}
\label{conditions}
\begin{cases} \partial_t P+\nabla \mathcal{E}=-eE_t/m, \\
V \times \left(\nabla \times V+eB/m \right)=- e E_t /m, 
\end{cases}
\end{equation}
where $E_t$ is the transversal electric field. 
 It is an electric field because, from $P=V+eA/m$ and $\mathcal{E}=V^2/2+e\Phi/m$, then
\begin{equation}
\label{Efield}
E=\frac{m}{e}\left(\nabla \frac{V^2}{2}+\partial_t V \right)+E_t.
\end{equation}
It is transversal being orthogonal to $V$: $V\cdot E_t=0$.
Any solution of (\ref{conditions}) is solution of motion. The simplest one is obtained setting $E_t=0$, or 
\begin{equation}
\label{simHam}
\partial_t P+\nabla \mathcal{E}=0.
\end{equation}
In such case \cite{footnote1}, the solution of (\ref{conditions}) satisfies the following fundamental equation:
\begin{equation}
\label{claudio}
\nabla \times V+\frac{eB}{m} =\frac{V}{\lambda}.
\end{equation}
The above equation has been already studied in another context, but only if coupled to a second equation for the modeling of two magnetofluids \cite{mahajan}, although the analogies are interesting,  they aren't considered here.\\
The physical meanings of $\lambda$, which is  
\begin{equation}
\label{lambda}
\lambda = \frac{V^2}{(e/m)V\cdot B+V\cdot \nabla \times V}.
\end{equation}
will be clarified in section~\ref{sec:sec5}.
 In this work, two kinds of solution of equation (\ref{claudio}) are mainly considered, the GP velocity and the gyro-particle velocity solutions, corresponding to $V/\lambda \sim (eB/m)$ and $\nabla \times V \sim-(eB/m)$, respectively. Before discussing the derivation of such kind of solutions, it is worth noticing that equation (\ref{claudio}) can be rewritten as
\begin{equation}
\label{claudioequiv}
V=\lambda \nabla \times P,
\end{equation} 
which means that a charged particle moves parallel to the \emph{curl} of the momentum $P$. In other words,  when  (\ref{simHam}) is verified, a moving charged particle ''creates" a magnetic field, $B$, satisfying (\ref{claudio}) or a vector potential, $A$, satisfying (\ref{claudioequiv}). Moreover, from \emph{Stoke}'s theorem, an integral interpretation can be given to (\ref{claudioequiv}). Suppose to close the orbit of the particle in the loop $\partial \Sigma$, where $\Sigma$ is any normal (smoothed and without holes) surface having $\partial \Sigma $ as boundary, then the flux of $V/ \lambda$ through $\Sigma$ is an action. Indeed,
\begin{equation}
\label{flux0}
\Phi_{V/\lambda}=\int_\Sigma \frac{V}{\lambda} \cdot   d^2 x =\oint_{\partial \Sigma} P \cdot V dt=\oint_{\partial \Sigma} (\mathcal{E}+L) dt,
\end{equation}
being $L$ the \emph{Lagrangian}. \\
In the remainder, the condition (\ref{simHam}) will be applied in (\ref{Efield}) for determining the \emph{effective} electric field:
\begin{equation}
E_\mathrm{eff}=\frac{m}{e}\left(\nabla \frac{V^2}{2}+\partial_t V \right).
\end{equation}
When a solution $V$ is found for $\dot V=(e/m)(E_\mathrm{eff}+V\times B)$, then the same solution satisfies  
\begin{equation}
\dot V= (e/m)\left[E_\mathrm{eff}+E_t+V\times( B+B_t)\right],
\end{equation}
if $E_t+V\times B_t=0$. Choosing $B_t=V\times E_t/V^2$, the system (\ref{conditions}) is rewritten as
\begin{equation}
\label{conditions2}
\begin{cases} \partial_t P+\nabla \mathcal{E}=-eE_t/m, \\
V \times \left[\nabla \times V+(e/m)(B-B_t) \right]=0,
\end{cases}
\end{equation}
and the solvability equation (\ref{claudio}) becomes
\begin{equation}
\label{claudio3}
\nabla \times V+\frac{e}{m} \frac{E_t\times V}{V^2}+\frac{eB}{m} =\frac{V}{\lambda}.
\end{equation}
As before, $\lambda$ is expressed by (\ref{lambda}), whilst $(e/m)E_t\times V/V^2$ is associated, as it will be shown in section~8, to a particular ''$E\times B$" \emph{drift} velocity.

\section{Guiding Particle velocity}
\label{sec:sec3}

 The GP solution of (\ref{claudio}) is characterized to be minimally coupled to $B$ through the \emph{Lorentz} force law being $V$ mostly parallel to it: $V\approx \lambda (e/m)B$. This implies that, for a GP, $|(eB/m)|>|\nabla \times V|$.  The GP velocity  is obtained substituting:
\begin{equation}
\nabla \times V= -i \mathbb{S} \cdot \nabla V,
\end{equation}
where $\mathbb{S}$ is a vector of $3\times3$ matrix components. In cartesian coordinates, $\mathbb{S}$ is  related to the antisymmetric \emph{Levi-Civita} symbol:
\begin{equation}
\{\mathbb{S}_{ik}\}_j=i\epsilon_{ijk}. 
\end{equation}
When equation (\ref{claudio}) can be inverted, the result is:
\begin{equation}
\label{formal0}
V=\left( 1+i \lambda \mathbb{S} \cdot \nabla \right)^{-1} \frac{\lambda eB}{m},
\end{equation}
being the differential operator $( 1+i \lambda \mathbb{S} \cdot \nabla)^{-1}$ defined as
\begin{equation}
\left( 1+i \lambda \mathbb{S} \cdot \nabla \right)^{-1} \equiv \sum_{n=0}^{\infty} (-i)^n \lambda^n (\mathbb{S} \cdot \nabla)^n.
\end{equation}
The equation (\ref{formal0}) means that the GP velocity belongs from $B$, after having  regularized it with $\lambda$, and propagating it to $V$. Such solution is the velocity of a particular  charged particle, the GP, in a magnetic field $B$ and in the effective electric field, consistently given  by
\begin{equation}
\label{GPE}
E_\mathrm{eff}=\frac{m}{e}\left(\nabla \frac{V^2}{2}+\partial_t V \right).
\end{equation}
It is possible to iteratively  obtain the GP solution, (\ref{formal0}), starting from the constant and uniform $B$ case. This means that also such approach leads to perturbative solutions \cite{footnote2}.
The leading solution is obtained when $B=|B| b^{(0)}$ is uniform, and $V=v_\| b^{(0)}$ is solution of (\ref{claudio}) for $\lambda=(m/e)v_\|/|B|$. When $B$ departs from being uniform the velocity $V^{(0)}$ is leaved to be 
\begin{equation}
V^{(0)}=\lambda^{(0)} \frac{e|B|}{m} b^{(0)}.
\end{equation}
The first order is obtained inserting the 0$^{th}$ solution into the LHS of (\ref{claudio}):
\begin{equation}
V^{(1)}=\lambda^{(1)}\frac{e|B|}{m} \left(  b^{(0)}+ \frac{m}{e|B|}\nabla \times V^{(0)}\right)=\lambda^{(1)}\frac{e|B|}{m}  \left[ b^{(0)}+ \frac{m}{e|B|}\nabla \times (\lambda^{(0)}\frac{e|B|}{m} b^{(0)})\right].
\end{equation}
For obtaining $\lambda^{(1)}$, which is a sort of \emph{regularization factor}, it is convenient introducing a measurable quantity, namely the parallel to $B$ velocity: $v_\| \equiv V \cdot b^{(0)}$. In such a way,
\begin{equation}
\lambda^{(0)}=\frac{mv_\|}{e|B|}
\end{equation}
and
\begin{equation}
\lambda^{(1)}=\frac{(mv_\|/e|B|)}{1+(m/e|B|) b^{(0)}\cdot \nabla \times (v_\|b)}.
\end{equation}
The first order solution is 
\begin{equation}
V^{(1)}=v_\| \frac{ b^{(0)}+(m/e|B|) \nabla \times (v_\| b^{(0)})}{1+(m/e|B|) b^{(0)}\cdot \nabla \times (v_\|b^{(0)})}.
\end{equation}
Such velocity is recognized as the GC velocity (for null \emph{magnetic moment}) if the GC transformation is truncated at first order \cite{white}.
 The velocity at all the orders is simply given by
\begin{equation}
\label{iter1}
V^{(n)}=v_\| \frac{b^{(0)}+(m/e|B|) \nabla \times V^{(n-1)}}{ 1+(m/e|B|) b^{(0)}\cdot \nabla \times V^{(n-1)}}.
\end{equation}
Instead of $v_\|=V\cdot b^{(0)}$, it can be used another kind of regularization:
the algebraic magnitude $|V|_s\equiv$ sgn$(V\cdot B)|V|= V \cdot b$. Note that the unit vector $b$ is parallel to $V$, whilst $b^{(n)}$ is parallel to $V^{(n)}$, written as
\begin{equation}
\label{iter2}
V^{(n)}=|V|_s \frac{ b^{(0)}+ (m/e|B|)\nabla \times V^{(n-1)}}{\left|  b^{(0)}+(m/e|B|)\nabla \times V^{(n-1)}\right|}=|V|_s b^{(n)}.
\end{equation}
Commonly,  $b$ denotes the unitary vector in the magnetic field direction, but, here, $b$ is parallel to $V$ while it is $b^{(0)}$ which is in the $B$ direction. The advantages of such choice will be clear in section~\ref{sec:sec8} and ~\ref{sec:sec9}.

\section{Gyro-particle motion}
\label{sec:sec4}

In this section it is considered another important solution of equation (\ref{claudio}) which applies when  $B\sim-(m/e)\nabla \times V$ ($\lambda \to \infty$). In this case, it looks reasonable taking the vector potential $A\sim-(m/e)V$ as solution. This cannot be done without conceptual difficulties. The vector potential $A$ has to be defined   in the same portion of space where $B$ lives and, furthermore, it is \emph{gauge} symmetric: $A\rightarrow A+\nabla g$ leaves $B$ unchanged.  When  $A\sim-(m/e)V$, both these properties seems broken but, only apparently.  An example is furnished by the constant and uniform magnetic field $B=B_0 e_z$. For such case, a standard choice of the vector potential is $A= r B \times e_r/2$, where the radial vector $re_r$ varies in the plane orthogonal to $B$, whilst $v_\perp =(e/m)\rho \times B$ being $\rho=\rho_L e_r$ varies on the circular ring of  \emph{Larmor} radius $\rho_L$. Specifically, $A=A(r,\gamma)$  and $v_\perp=v_\perp(\gamma)$, in poloidal coordinates. The magnetic field $B$ is obtained either from $B=\nabla \times A$ as from $B= -(m/e)\nabla \times v_\perp$. Moreover, at $r=\rho_L$ does exist a gauge function $g$ such that $A(\rho_L,\gamma)=-(m/e)v_\perp(\gamma)+\nabla g$: 
\begin{equation}
\label{gaugegamma}
g=(m/e)^2\mu \gamma,
\end{equation}
 being $\mu=(e/m)^2B_0\rho_L^2/2$ and $\nabla \mu =0$; moreover, $\gamma$ is the gyro-phase and $r \nabla \gamma=e_\gamma\equiv e_r \times e_z$. In this gauge, $v_\perp$ is a physical, measurable vector potential defined on a $1$D dimensional space, the ring $r=\rho_L$.  \\ For the general magnetic field case, the particle position is $x=X+\rho$, and the velocity is $v=\dot{X}+\dot{\rho}=\dot{\rho}=\sigma$, being $\dot X$ chosen to be zero. In such reference frame,  centered at $(X,\dot X)=(X,0)$,  the equation (\ref{claudio}) becomes 
\begin{equation}
\label{claudio2}
\nabla \times \sigma+\frac{eB(t,x)}{m}=\kappa \sigma,
\end{equation}
having replaced $V\rightarrow \sigma$, $X\rightarrow x$ and $\lambda \rightarrow 1/\kappa$. The gyro-particle is defined to be the one with $\nabla \times \sigma = -(e/m)B(t,x)$, or $\kappa = 0$, and whose orbit is closed.  As for the uniform and constant case, the gyrophase $\gamma \in S^1$ is the curvilinear coordinate describing such closed trajectory. The \emph{cyclotron} frequency is defined to be the time derivative of $\gamma$: $\dot \gamma \equiv \omega_c$, in this way $\sigma=\omega_c d \rho/d\gamma$. The ratio $|\sigma|/|\omega_c|$ is called \emph{Larmor} radius which becomes equivalent to $\rho_L=|d\rho/d\gamma|$.  Applying the \emph{Stoke}'s theorem for computing the magnetic flux $2\pi \Psi$ of $B$ through the normal (smoothed and without holes) surface $\Sigma$, whose boundary, $\partial \Sigma$ is the closed orbit traced by the gyrating particle, the following result is obtained:
\begin{eqnarray}
\nonumber
2\pi \Psi&=&\int_\Sigma B \cdot d^2 x=-(m/e)\oint_{\partial \Sigma} \sigma \cdot d\rho=-(m/e)\oint \omega_c  \left(\frac{d\rho}{d\gamma}\right)^2 d\gamma\\
\label{Psi0}
&=&-(m/e)\oint \omega_c \rho_L^2 d\gamma,
\end{eqnarray}
or 
\begin{equation}
2\pi \Psi=-\oint   \frac{\sigma^2}{(e\omega_c/m)} d\gamma.
\end{equation}
So that the magnetic flux through the gyro-particle orbit is expressed by the \emph{gyro-average} of $2\pi(m/e)\omega_c \rho_L^2$, or the \emph{gyro-average} of $2\pi(m/e)\sigma^2/\omega_c$. There is a particular set of coordinates where such motion is simplified. This happens when $\gamma$ expresses a local symmetry of $B$. It will be shown that, in such case, $\omega_c \rho_L^2$ is constant along the trajectory and it can be put outside the above integral which becomes: $\Psi=-(m/e)\omega_c\rho_L^2$.  

Respect to the general problem of a time varying and non homogeneous $B(t,x)$, it is possible to find a particular solution of (\ref{claudio2}) if expressed by
\begin{equation}
\label{sigma0}
 \sigma= \rho \times \Omega.
 \end{equation}
This solution is similar to the perpendicular velocity in (\ref{notation}), with $\rho$ and $\Omega$ to be determined. 
 The vector $\Omega=\Omega(t,x)$ is an \emph{angular velocity} vector field. 
The \emph{curl} of $\sigma$ is
\begin{equation}
\nabla \times ( \rho \times \Omega) = \rho \nabla \cdot \Omega -\rho \cdot \nabla \Omega + \Omega \cdot \nabla \rho - \Omega \nabla \cdot \rho.
\end{equation}
For the uniform magnetic field, $\Omega$ is given by $\Omega=(eB_0/m)e_z$, and only the fourth term survives. In the non homogeneous case it is required that $\nabla \cdot \Omega=0$ and $\nabla \cdot \rho=1$. The relation $\nabla \cdot \Omega=0$ means that  $\Omega$'s field lines  are closed (if not \emph{ergodic}), as for the magnetic field lines. The relation $\nabla \cdot \rho=1$ means that $\rho$ describes the spatial position of a 1D trajectory: $\rho=\rho(t,\gamma)$, so that $\nabla \cdot \rho =\nabla \gamma \cdot \partial_\gamma \rho=1$.  
Calling $L_\rho \Omega$ the \emph{Lie} derivative of $\Omega$ respect to $\rho$: $L_\rho \Omega =  \rho \cdot \nabla \Omega -\Omega \cdot \nabla \rho$, the \emph{curl} of $\sigma$ gives rise to the following relation, from (\ref{sigma0}) and (\ref{claudio2}) with $k=0$:
\begin{equation}
\label{eq41}
-\nabla \times \sigma=\Omega +L_\rho \Omega =\frac{eB}{m}.
\end{equation}
If (\ref{eq41}) is inverteble, \emph{e.g.} for sufficiently small gyro-radius, it could be obtained the solution
\begin{equation}
\label{Omega0}
\Omega=(1+L_\rho)^{-1} \frac{eB}{m},
\end{equation}
being
\begin{equation}
(1+L_\rho)^{-1}=\sum_{n=0}^\infty (-)^n L_\rho^n.
\end{equation}
It is worth noticing that adding to $\Omega$ a vector angular velocity along the direction $e_\rho$, the gyro-particle velocity in (\ref{sigma0}) doesn't change, so that it can be chosen $\Omega$ to be perpendicular to $\rho$.
Another interesting property of such solution is the following, being $\sigma= \dot \rho$, then
\begin{equation}
\frac{d}{dt}\frac{\rho^2}{2}=\rho \cdot \dot \rho =\rho \cdot \sigma =\rho \cdot \rho \times \Omega=0.
\end{equation}
The solution moves on the surface $S^2$ of a sphere whose radius can be identified with the  constant \emph{Larmor} radius $\rho_L$.
In this way, the magnetic flux linked to the closed orbit $2\pi \Psi$ becomes, from (\ref{Psi0}), proportional to the period it occurs to a particle for one revolution. In other words,  magnetic flux can be used as a time-like variable.\\
The same solution (\ref{Omega0}) can be obtained iteratively, starting from the uniform magnetic field but
stretching it a bit  and leaving it free to fluctuate.. In this way a recursive scheme is  found starting from the leading solution:
\begin{equation}
\sigma^{(0)}=\rho \times  \Omega^{(0)}=\rho \times (eB/m).
\end{equation} 
Applying the \emph{curl} to the above expression, it gives:
\begin{equation}
\label{zerogyrop}
\nabla \times \sigma^{(0)} =-eB/m  -\rho \cdot \nabla (eB/m) + (eB/m) \cdot \nabla \rho.
\end{equation} 
Multiplying (\ref{zerogyrop}) per $\sigma^{(0)}$, it is possible to define $\kappa^{(0)}$ as:
\begin{equation}
\kappa^{(0)}=\frac{\sigma^{(0)} \cdot \nabla \times \sigma^{(0)}}{|\sigma^{(0)}|^2}.
\end{equation} 
With the identity matrix $\mathbb{1}$, and the \emph{Lie} derivative $L_\rho (eB/m)$, the equation (\ref{zerogyrop}) is rewritten as
\begin{equation}
\label{zero2gyrop}
\nabla \times \sigma^{(0)} + eB/m  +L_\rho (eB/m) \cdot \left[ \mathbb{1}-\frac{\sigma^{(0)} \sigma^{(0)}}{|\sigma^{(0)}|^2} \right]= \kappa^{(0)} \sigma^{(0)}.
\end{equation} 
The velocity $\sigma^{(0)}$ is solution of motion only if $L_\rho (eB/m) \cdot  [|\sigma^{(0)}|^2\mathbb{1}-\sigma^{(0)} \sigma^{(0)}]=0$. This happens when $L_\rho (eB/m) = -k^{(0)} \sigma^{(0)}$, otherwise next order solution, $\sigma^{(1)}= \rho \times \Omega^{(1)}$, is processed. $\Omega^{(1)}$ can be written as
\begin{equation}
\Omega^{(1)}= \Omega^{(0)} - L_\rho \Omega^{(0)}=eB/m-\left[\rho \cdot \nabla (eB/m) - (eB/m) \cdot \nabla \rho \right].
\end{equation}  
The angular velocity field $\Omega^{(1)}$ is divergence free, being $L_\rho (eB/m)=-(eB/m)-\nabla \times \sigma^{(0)}$. The \emph{curl} of $\sigma^{(1)}$ becomes:
\begin{equation}
\label{firstgyrop}
\nabla \times \sigma^{(1)} = -eB/m+L_\rho (eB/m)-L_\rho \left[(eB/m-L_\rho (eB/m)\right]=-eB/m+L^2_\rho (eB/m)
\end{equation}
or
\begin{equation}
\nabla \times \sigma^{(1)} + eB/m-L^2_\rho (eB/m)\cdot \left[ \mathbb{1}-\frac{\sigma^{(1)} \sigma^{(1)}}{|\sigma^{(1)}|^2} \right]= \kappa^{(1)} \sigma^{(1)},
\end{equation} 
with, again,
\begin{equation}
\kappa^{(1)}=\frac{\sigma^{(1)} \cdot \left[\nabla \times \sigma^{(1)}+\Omega^{(0)}\right]}{|\sigma^{(1)}|^2}.
\end{equation} 
 A recursive scheme is realized:
\begin{equation}
\label{expansion}
\begin{cases}
\sigma^{(n)}=\rho \times \Omega^{(n)};\\
\Omega^{(n)}=\Omega^{(n-1)}+ (-1)^n L_\rho^n (eB/m)\\
\kappa^{(n)}=\sigma^{(n)} \cdot \left[\nabla \times \sigma^{(n)}+\Omega^{(n-1)}\right]/|\sigma^{(n)}|^2.
\end{cases}
\end{equation}
If exists $n$ such that $L^{(n)}_\rho (eB/m) \cdot  [|\sigma^{(n)}|^2\mathbb{1}-\sigma^{(n)} \sigma^{(n)}]=0$ then $\sigma^{(n)}$ is the velocity solution of motion. Otherwise, if $L^{(n)}_\rho (eB/m) \cdot  [|\sigma^{(n)}|^2\mathbb{1}-\sigma^{(n)} \sigma^{(n)}]\neq0$ for all $n$, and $|L^{n+1}_\rho (eB/m)|<|L^{n}_\rho (eB/m)|$, for all $n>n_0$, then the final angular velocity vector field, (\ref{Omega0}), is the sum of the geometrical series of $\Omega^{(n)}$. For such solution, $\kappa=0$ and the velocity is (\ref{sigma0}).
\\
In the former two sections the time dependency has been kept quite hidden. The problem is that what it has been defined to be GP or gyro-particle, can be useful on identifying particles only if it is considered a finite time window. Indeed it can happen that for $t\in \Delta t$ then $eB(t)/m \sim V/\lambda$, but, before or after a while, for $t^\prime \not\in \Delta t$ it is $eB(t^\prime)/m \sim -\nabla \times V$, and $V$ is not anymore a GP. In the following section we describe the generic particles in two particular reference frames that take explicitly care of the time dependency, the first is the gyro center reference frame $(X,V)$, the other is the GC reference frame that will be indicated by $(Q,U)$.
   
 \section{Charge motion in gyro-center and in GC coordinates}
\label{sec:sec5}

The GP and gyro-particle velocities are solutions of the \emph{Lorentz}'s equation and are obtained from (\ref{claudio}). It is possible to sum the velocities obtaining a general solution only if the magnetic field is constant and uniform, because such solutions live in orthogonal spaces, parallel and perpendicular to $B$. The aim of this  section is to analyze the possibility of adding the GP and the gyro-particle solutions to obtain the description of  motion for a generic charged particle, also for  a general magnetic field.
The vector position and velocity of the particle, at time $t$, are written as 
\begin{equation}
x = X+\rho,
\end{equation}
and
\begin{equation}
v = V+ \sigma,
\end{equation}
being $V=V(t,X)$ solution of (\ref{claudio}). The coordinates $(x,v)$ can be either GP coordinates, if $\rho = 0$ and $|\nabla \times V| < |eB/m|$, and/or gyro-particle coordinates, if $V=0$ and $\nabla \times \sigma \sim -eB/m$. As before mentioned, such identification can change with time: if $|\nabla \times V| < |eB/m|$ occurs at time $t=0$ but it isn't anymore true at a later time then the particle is well described as a GP only for a finite time interval near $t=0\in \Delta t$. \\
In this present work the gyro-center is defined by the following statement: the phase space frame of reference where the particle orbit makes a closed loop is the gyro-center reference frame \cite{footnote3} and indicated by $(X,V)$: gyro center position $X$  and gyro center velocity $V$. It is here assumed that $V\mid_{t=0}=U$ is solution of (\ref{claudio}) and, moreover, it is a GP if $\rho \mid_{t=0}=0$, with position indicated by $X\mid_{t=0}=Q$.  The GC coordinates are the coordinates $(Q,U)$, and they are defined as the gyro-center coordinates computed at time $t=0$, if it is known that
\begin{equation}
\label{GCV}
U=\left( 1+i \lambda_0 \mathbb{S} \cdot \nabla \right)^{-1} \frac{\lambda_0 eB_0}{m}
\end{equation}
 being $\lambda_0=\lambda(0,Q)$ and $B_0=B(0,Q)$, the $\lambda$ length and the magnetic field, respectively, computed at $t=0$.
Such magnetic field, $B_0$, is called \emph{equilibrium} magnetic field because it is frozen at $t=0$: $\partial_t B_0=0$. The GC coordinates, if $\rho\mid_{t=0}=0$, describe the GP moving in a equilibrium magnetic field.
\\
The gyro center velocity is not generally known as it is the GC velocity. The difference between $X$ and $Q$ is called the GC \emph{displacement} and it is indicated with $\xi$. The displacement vector depends on the time variation of the magnetic field from $B_0=B(0,Q)$ to $B(t,X)$, which is called the magnetic field fluctuation: $\delta_\xi B= B(t,X)-B(0,Q)$. It is also useful introducing the perturbed velocity, $\delta_\xi U=V-U$, computed by
\begin{equation}
\dot \xi = \delta_\xi U.
\end{equation}
 In this way the relation between gyro center and GC coordinates is made explicit:
\begin{equation}
X=Q+\xi,
\end{equation}
\begin{equation}
V=U+\delta_\xi U.
\end{equation}
Also the particle coordinates $(x,v)$ can be expressed respect to the GC reference frame:
\begin{equation}
x=Q+\xi+\rho,
\end{equation}
\begin{equation}
v=U+\delta_\xi U+\sigma.
\end{equation}
It is worth noticing that only $Q$ and $U$ can be explicitly obtained starting from  the \emph{equilibrium} magnetic field $B_0$. Many kinds of particle motion can be realized with the above description, and it can be helpful consider an orderings between such coordinates. It is like for the hands of an analog clock, the hour hand makes a revolution with a slow frequency $\omega_e$, minutes are specified by an intermediate frequency, $\omega_d$, and seconds by the highest frequency, $\omega_c$, with $\omega_c > \omega_d > \omega_e$. The charge motion is a composite motion in some way similar to the hands motion in an analog clock. To specify what time is now, you first look at the hour hand, then, with some approximation, also at the minutes hand, and rarely also at the seconds hand. The same is for the charged particle: the GC coordinates \emph{can be}, like hours, the first, rough, representative of the particle coordinates (computed at t=0), the displacement and the perturbed velocity \emph{can be}, like minutes, very useful if summed to the GC coordinates, and a finer motion description is obtained by the gyro-center coordinates;  the \emph{Larmor} radius and the $\sigma$ velocity, \emph{can be} ignored, like seconds. Obviously, this is just an example because there are other situations where only the evolution of $\rho$ and $\sigma$ are considered relevant. 

Concerning the gyro-center coordinates, the particle motion can be easily described starting from the time derivative of $V$, which satisfies the following  \emph{Lorentz} force law: 
\begin{equation}
\dot V =\frac{e}{m} (E_\mathrm{eff}+V\times B)=\nabla \frac{V^2}{2}+\partial_t V+V\times B,
\end{equation}
being $\partial_t P +\nabla \mathcal{E}=0$, as  seen in (\ref{conditions}).

The energy $\varepsilon$ of the particle is
\begin{equation}
\label{energia}
\varepsilon=\frac{V^2}{2}+\frac{e}{m}\Phi(t,X) + \frac{\sigma^2}{2}+\sigma \cdot V + \frac{e}{m}\delta_\rho \Phi=\mathcal{E}+ \frac{ \sigma^2}{2}+ \sigma \cdot V + \frac{e}{m}\delta_\rho \Phi
\end{equation}
being $\delta_\rho$ the incremental operator, defined by $\delta_\rho O(t,X+\rho)=O(t,X+\rho)-O(t,X)$ for a general field $O$. If not explicitly said, the gyro-center velocity  $V$, the electro-magnetic field $E$ and/or $B$, and the related potentials $\Phi$ and $A$, are referred to the time $t$ and position $X$.  The momentum of the particle, $p$, is the sum of $P=V+(e/m)A$ and $\pi$, defined as $\pi=\sigma+\delta_\rho A$. Explicitly:
\begin{equation}
p=V+\sigma +\frac{e}{m}A +\frac{e}{m}\delta_\rho A=P+\pi.
\end{equation} 
The \emph{Lorentz} force law
\begin{equation}
\dot{v}=\frac{e}{m}\left[ E(t,x) +v\times B(t,x)\right]
\end{equation}
is rewritten as
\begin{equation}
\dot \sigma= \frac{e}{m}\left[\delta_\rho E +V \times \delta_\rho B\right] +\sigma \times \frac{e}{m}(B+\delta_\rho B)
\end{equation}
It is worth noticing that $\sigma$ can be interpreted as the velocity of a particle which is weakly coupled to the ''electric field":$\delta_\rho E +V \times \delta_\rho B$, that  goes to \emph{zero} together with $\rho$.
The latter equation has to be compared with the time derivative of $\sigma$, being $\sigma=v(t,x)-V(t,X)$, so that 
\begin{equation}
\dot \sigma= \partial_t \sigma +v\cdot \nabla v -V\cdot \nabla V=\partial_t \sigma +V\cdot \nabla \sigma +\sigma \cdot \nabla V +\sigma \cdot \nabla \sigma.
\end{equation}
From the vectorial identity (\ref{vectId0})
applied for $\mathrm{a}=\sigma$ and $\mathrm{b}=V$ (and also for $\mathrm{a}=\mathrm{b}=\sigma$), the \emph{Lorentz} force law becomes
\begin{equation}
\label{sigmaevol}
 \partial_t p + \nabla \varepsilon= V\times \left[ \nabla \times \sigma + \frac{e}{m} \delta_\rho B \right]+  \sigma \times \left[ \nabla \times \sigma +\nabla \times V + \frac{e}{m} B +\frac{e}{m} \delta_\rho B \right].
\end{equation}
Some of the solutions of the former equation can be chosen to satisfy the following system of equations \cite{footnote4}:
\begin{equation}
\label{totalsystem}
\begin{cases} \partial_t p+\nabla \varepsilon=0, \\
\nabla \times V+eB/m=V/\lambda,\\
 \nabla \times \sigma+V/\lambda+ e\delta_\rho B/m=0.
\end{cases}
\end{equation}
The velocity  $\sigma$ is $\sigma=\rho \times \Omega(t,X+\rho)$, as in (\ref{sigma0}), even if it it is not anymore a gyro-particle velocity.  In this case, $\sigma$ is the particle velocity expressed in the gyro-center reference frame $(X,V)$ at time $t$. It is worth to note that the considered solution is particular being $\nabla \times v=\nabla \times V+\nabla \times \sigma=-(e/m)B-(e/m)\delta_\rho B=-(e/m)B(t,x)$\cite{footnoten}.
 The angular velocity $\Omega(t,X+ \rho)$ is obtained from (\ref{Omega0}) with the substitution $eB/m \rightarrow V/\lambda$. Explicitly:
\begin{equation}
\Omega(t,X+\rho)=(1+L_\rho)^{-1} \left( \frac{V}{\lambda}+ \frac{e\delta_\rho B}{m} \right).
\end{equation}
The angular velocity field can be conveniently rewritten as:
\begin{equation}
\Omega(t,X+ \rho)=V/\lambda +\delta_\rho \Omega.
\end{equation}
It is possible to give a physical meaning to $\omega_c$ and $\lambda$ identifying the \emph{cyclotron} frequency for general $B$ to be the algebraic magnitude of $\Omega(t,x)$: 
\begin{equation}
\label{assumption}
\Omega(t,x)=\omega_c(t,x) b(t,x).
\end{equation}
If the \emph{transit} frequency, $\omega_t$, is defined to be the limit behavior of $\omega_c$ for $\rho \to 0$:
\begin{equation}
\omega_t(t,X) \equiv \lim_{\rho \to 0} \omega_c(t,x),
\end{equation}
then $\lambda$ can be obtained from:
\begin{equation}
\label{assumption2}
\frac{V(t,X)}{\lambda(t,X)}= \lim_{\rho \to 0} \Omega(t,x)=\omega_t(t,X) b(t,X).
\end{equation}
The pitch-length $\lambda$ is the path length ran by GP in the time lapse $\omega_c^{-1}(t,X)$.
 In this way it is re-obtained the known property that gyrokinetics is able to describe multi-scale  systems. Moreover, if $|\nabla \times V| <|eB/m|$ the velocity field $V$ can be chosen to be 
\begin{equation}
\label{gyroV0}
V=\left( 1+i \lambda \mathbb{S} \cdot \nabla \right)^{-1} \frac{\lambda eB}{m},
\end{equation}
as in  (\ref{formal0}). In this case the gyro-center velocity $V$ is mostly along the magnetic field direction whilst $\sigma$ is mostly orthogonal to $B$, so that $\lambda$ can be considered the longitudinal scale of the problem, whilst $\rho_L$ the perpendicular one. Commonly, both the scales are relevant and the \emph{cyclotron} frequency determines the ratio $\omega_c(t,x)=|\sigma|/\rho_L$ and $\omega_c(t,x)\to \omega_t(t,X)=|V|_s/\lambda$ for $\rho_L \to 0$. \\
 The problem with gyro-center velocity is that rarely the solution (\ref{gyroV0}) can be considered correct  for a large time interval, this happens only for \emph{slowly varying} magnetic fields. However, the system of equations (\ref{totalsystem}) can be expressed by the GC coordinates instead of the gyro center ones:
 \begin{equation}
\label{totalsystemGC}
\begin{cases} \partial_t p+\nabla \varepsilon=0, \\
\nabla \times \delta_\xi U+kU+e \delta_\xi B/m=(1-\lambda_0k)\delta_\xi U/\lambda_0,\\
 \nabla \times \sigma+(1-k\lambda_0)(U+\delta_\xi U)/\lambda_0+ e\delta_\rho B/m=0,
\end{cases}
\end{equation}
where $k=1/\lambda-1/\lambda_0$, and $U$ is already known from (\ref{GCV}). It is worth noticing that the equation for the perturbed velocity $\delta_\xi U$, as for $\sigma$ and $U$, is always the same fundamental equation (\ref{claudio}) applied for different ''magnetic" fields, and it is possible to make a group containing all solutions of (\ref{claudio}). Moreover, it is interesting to notice the hierarchy between solutions: $U$ satisfies (\ref{claudio}) for $B\to B_0$ and $\lambda \to \lambda_0$, $\delta_\xi U$ satisfies the same equation for $B \to (m/e)k U+\delta_\xi B$ and $\lambda \to \lambda_0/(1-k \lambda_0 )$, or, in other words, the solution for $\delta_\xi U$ is slave of the solution for $U$ and $\lambda_0$. Once obtained the gyro center velocity $V=U+\delta_\xi U$, the same remark is true for $\sigma$, which is solution of (\ref{claudio}) for $B \to (m/e)V/\lambda$, or, $\sigma$ is slave of $V$ and $\lambda$ solutions which, in turn,  are slave of $U$ and $\lambda_0$ solutions, which in turn, depends on $B_0$.
 
In summary, the considered particle is represented in gyro-center coordinates by 
 \begin{equation}
 \label{chargeCoord}
x = X+\rho,
\end{equation}
 \begin{equation}
 \label{chargeCoord2}
 v=V+\rho \times \Omega=V+\rho \times V/\lambda + \rho \times \delta_\rho \Omega,
 \end{equation}    
  where $(X,V)$ are computed  at each time $t$ in the gyro-center description or substituted to $X=Q+\xi$ and $V=U+\delta_\xi U$ in the GC description.
  
  Up to know it has been considered general magnetic fields and two interesting solutions for the particle motion, in the next section it is considered a particular magnetic field, often encountered in the description of laboratory and astrophysical plasmas, and the corresponding general solution for  the particle motion.

\section{Charge motion in an axisymmetric magnetic field}
\label{sec:sec6}

In this section the charged particle motion is considered when the magnetic field shows some symmetries. If the system is symmetric in the toroidal $\phi$ direction, the magnetic field can be written as 
\begin{equation}
\label{hybrid}
B=\nabla \psi \times \nabla \phi + F \nabla \phi,
\end{equation}
 being $\phi$  the toroidal angle, $\nabla \phi =e_\phi/R$ ($e_\phi$ the unitary vector in the $\phi$ direction and $R$ the major radius in cylindrical coordinates), $\psi$ the poloidal magnetic flux. Moreover, $\nabla \psi \cdot \nabla \phi =0$ and $\partial_\phi F=0$, for the considered symmetry. This representation, mostly used for describing tokamak magnetic fields, is very similar to the one used on describing the velocity in equation (\ref{notation}). Indeed, it represents a magnetic field made by helicoidal magnetic field lines and can be decomposed into the sum of a toroidal field, $B_t$, whose direction is along the magnetic axis, $e_\phi$, and a poloidal  field, $B_p=\pm|B_p| e_\theta$, for a poloidal angle $\theta$ in toroidal coordinates. Magnetic field lines can be thought as proceeding towards the magnetic axis $e_\phi$ and simultaneously winding around it.  Such helicoidal magnetic field lines are characterized by the \emph{winding number} $\mathrm{i}_B=q^{-1}$, being   $q$  the \emph{safety factor}.
 
  For a closed magnetic field line,  $n$ toroidal cycles match $m$  poloidal ones, so that $q=m/n$ is a rational number, $q\in \mathbb{Q}$.  There is a \emph{flux coordinates} system where such description is easily visualized because the magnetic lines are mapped into straight lines. The surface made of magnetic field lines corresponding to a given $q$ is identified by the value of $\psi$, proportional to the \emph{poloidal magnetic flux} $\phi_p=2\pi \psi$. If $q \in \mathbb{Q}$ then $\psi$ is said to be \emph{resonant}. It is useful introducing a (toroidal) radial coordinate which labels $\psi$, in this way $r=r(\psi)$ and $q=q(r)$. The radius  $r$ depends on the intensity of magnetic field, for this reason, $r$ is considered a generalized radial coordinate. It is also introduced a generalized poloidal angle, $\vartheta \in S^1$, so that:
  \begin{equation}
 \label{sf}
 \frac{1}{q(r)}\equiv \frac{B\cdot\nabla \vartheta}{B\cdot\nabla\phi}.
 \end{equation}

 In the straight flux coordinates the magnetic field is expressed by 
 \begin{equation}
 \label{clebshA}
 B=\nabla \psi \times \nabla (\phi -q \vartheta).
 \end{equation}
  From (\ref{clebshA}), the vector potential is immediately found:
 \begin{equation}
 \label{Aclebsh}
 A=\psi\nabla \phi +q\vartheta \nabla \psi +\nabla g.
 \end{equation}  
 Thus vector potential representation is multivalued. However, it can be shown that  $2\pi q \nabla\psi$ is the gradient of the toroidal magnetic flux $\phi_t=-2\pi \Psi$ that becomes the $gauge$ function that leaves $B$ single valued. Moreover, such representation shows that $\psi$ can be defined through the toroidal component of $A$: $\psi \equiv |\nabla \phi|^{-1} (A-\nabla g)\cdot e_\phi$. A further simplification on fields mapping is obtained with the following \emph{Clebsh} potentials: $\gamma$ and$\Psi$. They are defined as $\gamma=\vartheta- \phi/q$ and from $\nabla \Psi=-q\nabla \psi$. The magnetic field can then be written in the \emph{Clebsh} representation as
 \begin{equation}
 \label{clebsh}
 B=\nabla \Psi \times \nabla \gamma,
 \end{equation}
 while the vector potential becomes:
 \begin{equation}
 A=\Psi \nabla \gamma +\nabla g,
 \end{equation}
  and $\Psi$ can be defined as $\Psi \equiv  |\nabla \gamma|^{-1} (A-\nabla g)\cdot e_\gamma$. Note that the \emph{Clebsh} representation can be adopted also when $B$ has no symmetries \cite{fluxcoordinates}.
  \\
 Armed with the above magnetic field representations it is now possible to address and, in some circumstances, to resolve the general case of axisymmetric system.
Firstly, equation (\ref{hybrid}), for representing $B$, is used  into the equation of motion (\ref{claudio}).  
The velocity $v$, solution of (\ref{claudio}) is decomposed in the sum $v=v_p+v_0$. The particular solution, $v_p$, satisfies 
\begin{equation}
\label{particular}
\nabla \times v_p+ \frac{eB}{m}=  \frac{v_p}{ \lambda_p},
\end{equation}
 while $v_0$ is the homogeneous solution satisfying:
\begin{equation}
\label{curlSpec}
\nabla \times v_0= \frac{ v_0}{\lambda_p}.
\end{equation}
Note that (\ref{curlSpec}) is known as the \emph{curl spectrum equation}, or as the \emph{force-free  equation} \cite{lust54,chandra57}. $v_0$ is often called a \emph{Beltrami} field \cite{beltrami}. It has been used in \cite{Woltjer} to find a magnetic field with fixed helicity and minimum energy for the description of some kind of astrophysical sources. 
As for example, it is reported below the solution find by \cite{Cantarella} for the \emph{spheromak}, which is the general solution of equation (\ref{curlSpec}), when $\lambda$ is \emph{constant}, in spherical boundary domain, $e_r\cdot v_0=0$, for $e_r$ orthogonal to the ball of radius $a$. Writing $v_0=u_r e_r +u_\theta e_\theta + u_\phi e_\phi$ in  spherical coordinates $(r,\theta,\phi)$, then
\begin{equation}
u_r=(l^2 -\frac{1}{4})r^{-\frac{3}{4}}J_l(r/\lambda)P^m_n(\cos \theta) {\cos m\phi \choose \sin m\phi} ,
\end{equation}
\begin{equation}
u_\theta= \frac{1}{r} \partial_r [\sqrt{r}J_l(r/\lambda)] \partial_\theta P^m_n(\cos \theta) {\cos m\phi \choose \sin m\phi} \mp  \frac{ m}{\lambda\sqrt{r}}J_l(r/\lambda)\frac{P^m_n(\cos \theta)}{\sin \theta} {\sin m\phi \choose \cos m\phi},
\end{equation}
\begin{equation}
u_\phi=\mp \frac{m}{r} \partial_r [\sqrt{r}J_l(r/\lambda)] \frac{P^m_n(\cos \theta)}{\sin \theta} {\sin m\phi \choose \cos m\phi} -\frac{m}{\lambda \sqrt{r}}J_l(r/\lambda) \partial_\theta P^m_n(\cos \theta) {\cos m\phi \choose \sin m\phi},
\end{equation}
 with $l=n+1/2$, $m$ (not to be confused with the mass) is $m \in \{0,1,...,n\}$ and 
  $J_l$ are \emph{Bessel} functions (of first kind, see (\ref{bessel})) and $P^m_l$  are associated \emph{Legendre} functions.

Coming back to the toroidal symmetric system, $v_p$ and $\lambda_p$ can be easily chosen to be:
\begin{equation}
\label{psol}
v_p= -(e/m) \psi \nabla \phi \qquad \mbox{               and              } \quad \lambda_p=-\psi/F.
\end{equation}
Such particular solution is proportional to the toroidal component of $A$: $v_p=-(e/m)A_\phi e_\phi$, see  (\ref{Aclebsh}) with  $\partial_\phi g=0$. 
The \emph{Beltrami} field $v_0$ is obtained starting form the following representation (to be compared with the velocity representation in (\ref{notation})):
\begin{equation}
\label{v_0}
v_0=\nu_0 [\nabla \mathcal{P}_\phi \times \nabla \phi + H_0 \nabla \phi],
\end{equation}
where the quantity in parenthesis is a special magnetic field with the poloidal magnetic flux $2\pi \mathcal{P}_\phi$ and a toroidal component $H_0 /R$, being $|\nabla \phi|=R^{-1}$. Moreover, for preserving the toroidal symmetry $\partial_\phi \mathcal{P}_\phi=\partial_\phi H_0=\partial_\phi \nu_0=0$. The \emph{curl} of (\ref{v_0}) gives:
\begin{equation}
\nabla \times v_0= \nabla \nu_0 \times ( \nabla \mathcal{P}_\phi \times \nabla \phi)+\nu_0 \nabla \times \nabla \mathcal{P}_\phi \times \nabla \phi+ \nabla (\nu_0 H_0) \times \nabla \phi.
\end{equation} 
Being
$$\nabla \nu_0 \times ( \nabla  \mathcal{P}_\phi \times \nabla \phi )=-(\nabla \nu_0 \cdot \nabla  \mathcal{P}_\phi) \nabla \phi
$$
and 
$$
\nabla \times( \nabla P_\phi \times \nabla \phi) = -R^2 \nabla \phi \ \nabla \cdot \frac{\nabla P_\phi}{R^2}=-\nabla \phi \Delta^\star P_\phi,
$$

where $\Delta^\star$ is the \emph{Shafranov} differential operator. 
The equation (\ref{claudio}) becomes:
\begin {equation}
-(\nabla \nu_0 \cdot \nabla  \mathcal{P}_\phi) \nabla \phi-\nabla \phi \Delta^\star P_\phi+\nabla (\nu_0 H_0) \times \nabla \phi=-\frac{F\nu_0}{\psi}[\nabla \mathcal{P}_\phi \times \nabla \phi + H_0 \nabla \phi].
\end{equation}
 Comparing the toroidal components:
 \begin{equation}
 \label{toro1}
 \nabla \nu_0 \cdot \nabla  \mathcal{P}_\phi+\Delta^\star P_\phi=\frac{F\nu_0 H_0}{\psi}
 \end{equation}
 and the poloidal ones:
\begin{equation}
\label{polo1}
 \nabla (\nu_0 H_0) \times \nabla \phi=-\frac{F\nu_0}{\psi}\nabla \mathcal{P}_\phi \times \nabla \phi,
 \end{equation}
 The following solution for $\nu_0$ and  $H_0$, are obtained:
  \begin{equation}
 \nu_0=-\frac{e\psi}{mF}=(e/m) \lambda_p  \quad \mbox{ and } \qquad
 H_0=-\frac{F \mathcal{P}_\phi}{\psi}=\mathcal{P}_\phi/\lambda_p.
 \end{equation} 
 The total velocity is
 \begin{equation}
 \label{symmsol}
 v=v_p+v_0=-\frac{e\psi}{mF} \nabla \mathcal{P}_\phi \times \nabla \phi+\frac{e}{m}\left(\mathcal{P}_\phi - \psi \right) \nabla \phi.
 \end{equation}
 The toroidal component of $v$ clarifies the meaning of $\mathcal{P}_\phi$, being proportional to the toroidal momentum $p_\phi=e_z\cdot x\times p=e_z \cdot x \times \left[v + (e/m) \times A \right]$:
 \begin{equation}
 \mathcal{P}_\phi=\psi+\frac{m}{e}R v_\phi \equiv m p_\phi/e,
 \end{equation}
 In an axisymmetric (and isolated) system $p_\phi$ is a constant of motion, which is immediately verifyed for $\mathcal{P}_\phi$, being
 \begin{equation}
  v\cdot \nabla \mathcal{P}_\phi=0,
 \end{equation} 
directly from the chosen representation (\ref{v_0}) of $v_0$. The velocity is expressed through a special magnetic field $(m\omega_t/e) b$. Indeed, $v=(m/e)\omega_t\lambda_pb$, coherently with equation (\ref{assumption2}), being $b$  parallel to $v$ and $(m/e)\omega_t b= \nabla \mathcal{P}_\phi \times \nabla \phi+\mathcal{F}\nabla \phi$, with $\mathcal{F}=F+\mathcal{P}_\phi/\lambda_p=F(1-\mathcal{P}_\phi/\psi)$. As done for the magnetic field, the velocity field lines lie on the magnetic flux surface $\mathcal{P}_\phi$ and a velocity \emph{winding number}, $\mathrm{i}_v$ is defined to be 
\begin{equation}
\mathrm{i}_v\equiv \frac{v\cdot\nabla \vartheta}{v\cdot\nabla\phi},
\end{equation}
analogously to the inverse of the safety factor.
When $\mathrm{i}_v \in \mathbb{Q}$ then $\mathcal{P}_\phi$ is called \emph{resonant} toroidal momentum and particles trace closed trajectories. It is worth noticing that for closed orbit the linked flux of $v/\lambda_p$ is related to the integral in (\ref{flux0}). Moreover, if the orbit is closed the frequency $\omega_d$ to make a cycle in $\phi$ is commensurable with the frequency $\omega_b$ to make a cycle in $\vartheta$, and both are multiple of $\omega_\mathrm{res}$, the characteristic frequency, or \emph{resonant} frequency, of the closed orbit. Moreover, for the closed orbit, $\mathfrak{i}_v=\omega_b/\omega_d\in \mathbb{Q}$ and $\omega_\mathrm{res}, \omega_d$ and $\omega_b$ are linearly dependent over $\mathbb{Q}$.  Analogously to the resonant $\psi$, the resonant $\mathcal{P}_\phi$ plays a fundamental role in the gyro-kinetic dynamics as it can be seen in the analysis of \emph{Poincar\'{e}} maps studied in \cite{briguglio2014}.

The toroidal momentum in flux unit, $\mathcal{P}_\phi$, can be obtained, once $F$ is known, from the explicit solution of equation (\ref{toro1}), here concisely rewritten:  
 \begin{equation}
 \label{diffPphi}
 \lambda_p R^2 \nabla \cdot \left( \frac{\nabla \mathcal{P}_\phi}{\lambda_p R^2} \right) \equiv\Delta_{\lambda_p R^2} \mathcal{P}_\phi=-\mathcal{P}_\phi,
 \end{equation} 
 being $\lambda_p=-\psi/F$
 with the operator $\Delta_S \equiv S \nabla \cdot S^{-1}\nabla$ equivalent to $\Delta^\star$ if $S\propto R^2$, which means that  $\lambda_p$ is a constant, or $F\propto \psi$. 
  
  Two solutions for $\mathcal{P}_\phi$, without solving equation (\ref{diffPphi}), are described here, for the GP velocity $V$, and  in the next section, for the gyro-particle velocity, $\sigma$. From equation (\ref{iter1}), the GP toroidal momentum is
 \begin{equation}
 \mathcal{P}^{(n)}_\phi=\psi+  \frac{1+ (m/e)(R/F)e_\phi \cdot \nabla \times V^{(n-1)}}{ 1+(m/e)(b^{(0)}/|B|)\cdot \nabla \times V^{(n-1)}}\frac{mFv_\|}{e|B|},
 \end{equation} 
 so that $\mathcal{P}^{(0)}_\phi=\psi+(m/e)Fv_\|/|B|$. Otherwise, from equation (\ref{iter2}), $\mathcal{P}_\phi$ becomes 
  \begin{equation}
 \mathcal{P}^{(n)}_\phi=\psi+  \frac{1+ (m/e)(R/|B|) e_\phi \cdot \nabla \times V^{(n-1)}}{ |1+(m/e)(b^{(0)}/|B|)\cdot \nabla \times V^{(n-1)}|}\frac{mF|V|_s}{e|B|},
 \end{equation} 
  with the new regularization schema proposed ($|V|_s$ instead of $v_\|$). 
 For the GP velocity  $V\sim-(e/m)\psi B/F$, the velocity (\ref{symmsol}) can be approximated to 
\begin{equation}
-\frac{e\psi}{mF} \nabla \mathcal{P}_\phi \times \nabla \phi+\frac{e}{m}\left(\mathcal{P}_\phi - \psi \right) \nabla \phi \sim -\frac{e\psi}{mF} \nabla \psi \times \nabla \phi-\frac{e\psi}{m} \nabla \phi,
\end{equation}
which means that $\mathcal{P}_\phi\sim 0$ and, $\nabla \mathcal{P}_\phi \sim \nabla \psi$ or $F\rightarrow \infty$.
Indeed, the GP velocity is mainly directed towards $B$. The \emph{drift velocity} expresses the deviation of the GP  from $\lambda (e/m) B$. In the present work, the \emph{drift velocity} is indicated with $V_D$ and defined as: 
\begin{equation}
\label{driftdef}
V_D \equiv \lambda  \nabla \times V=-\frac{e\psi}{mF} \nabla(\mathcal{P}_\phi-\psi)\times \nabla \phi +\frac{e\mathcal{P}_\phi}{m} \nabla \phi,
\end{equation} 
where the last equivalence is obtained for $\lambda =\lambda_p=-\psi/F$ and $V=(e/m)\lambda_p \nabla \mathcal{P}_\phi \times \nabla \phi+(e/m)(\mathcal{P}_\phi - \psi) \nabla \phi$. It is worth noticing that $\omega_V=\nabla\times V$ is sometimes called the \emph{vorticity}, so that $V_D=\lambda_p \omega_V$.\\
 It can be consistently checked that $V_D$ is smaller than $V$ when $\mathcal{P}_\phi\sim 0$ and, $\nabla \mathcal{P}_\phi \sim \nabla \psi$ or $F\rightarrow \infty$. It is worth noticing that the last condition means that the magnetic field is mostly toroidal, as it happens in tokamaks where GPs are believed to have an important role. \\
From the definition of drift velocity, it is also possible to obtain another useful expression of $\mathcal{P}_\phi$, multiplying both sides of (\ref{driftdef}) per $(m/e)Re_\phi$ we obtain:
\begin{equation}
\mathcal{P}_\phi=\frac{m\lambda_p R}{e}e_\phi \cdot \nabla \times  V=-\frac{m\psi}{eF}R^2\nabla \phi \cdot \nabla \times V.
\end{equation}
The last equation, together with (\ref{diffPphi}) allows to express the gradient of $\mathcal{P}_\phi$ for the general case ($V\to v$, which is not anymore the GP velocity):
\begin{equation}
\nabla \mathcal{P}_\phi=-\frac{m\lambda_p R^2}{e} v\times \nabla \phi.
\end{equation}

 \section{Gyro-symmetry}
\label{sec:sec7}

The conserved momentum for the gyro-particle
 velocity $\sigma$, when $\dot X=0$ and $\nabla \times \sigma=-(e/m)B$, needs another treatment. 
The representation (\ref{clebsh}) is better conceived for the present case because it describes general   magnetic configurations and, specifically, also not symmetric magnetic fields. The particular solution of (\ref{claudio2}), like in (\ref{psol}), is 
\begin{equation}
\sigma_p=-(e/m)\Psi \nabla \gamma \qquad \mbox{ for } \qquad \lambda_p \to \infty \,\,\,\mbox{( or }\kappa_p=0\mbox{)}.
\end{equation}
The homogeneous equation is reduced to $\nabla \times \sigma_0=0$ with the solution $\sigma_0=-(e/m)\nabla g$ so that 
\begin{equation}
\label{ggauge}
\sigma=-(e/m)(A+\nabla g),
\end{equation}
This result was already shown for the uniform magnetic field case. As in (\ref{gaugegamma}), a  good $gauge$ choice is the following:
\begin{equation}
 g=(m/e)^2 \mu \gamma,
 \end{equation}
 if  $\mu$ is constant: $\nabla \mu=0$.
 The gyro-particle velocity becomes 
 \begin{equation}
 \label{gyrosol2}
 \sigma=-(m/e)\left[\mu+(e/m)^2\Psi\right] \nabla \gamma.
 \end{equation}
 The same expression, (\ref{gyrosol2}),  just obtained without requiring any symmetry, can be also deduced from (\ref{symmsol}), when a symmetry is evident. With the following substitution: $\phi \to \gamma$, $\psi \to \Psi$ and $\mathcal{P}_\phi \to -(m/e)^2 \mu$, the velocity is 
  \begin{equation}
 \label{symmsol2}
 \sigma=-\frac{m\Psi}{eF} \nabla \mu \times \nabla \gamma-\frac{m}{e}\left[\mu + (e/m)^2\Psi \right] \nabla \gamma;
 \end{equation}
for the gyro-particle $\lambda_p \to \infty$ or $F \to 0$, so that  $\nabla \mu =0$ must be imposed and the relation (\ref{gyrosol2}) re-obtained but, now, $\gamma$ is an ignorable variable expressing the symmetry of the system and $\mu$ is the constant of motion associated to such symmetry.  This means that the \emph{gyro-symmetry} in $\gamma$ was also present, but hidden, in the first derivation of (\ref{gyrosol2}).\\
 In analogy to $\mathcal{P}_\phi$, the magnetic moment is defined as $(m/e)\mu = -b \cdot \rho \times \left[\sigma+(e/m)A \right]$, for the unit vector $b=\Omega/\omega_c$, where the angular velocity vector $\Omega$ is defined in (\ref{Omega0}) and, for the symmetric case, $\partial_\gamma \omega_c=0$. This definition of the magnetic moment for the gyro-particle is consistent with (\ref{gyrosol2}) if $\nabla \gamma \cdot b \times \rho=1$, or $\nabla \gamma=e_\gamma/\rho_L$ and $e_\gamma, b$ and $e_\rho$ define a tern of orthogonal unit vectors basis. \\
 The gyro-symmetry is important because is linked to the property of the gyro-particle to be closed, that is by definition in the present approach. In other words, when there is a closed orbit, it is possible to define a magnetic moment, a gyro-angle and a gauge, product of the two times $(m/e)^2$, so that such particle feels the force on itself independently from the gyro-angle. In this way it is assumed possible to have $\omega_c=\omega_c(t,X,\rho_L)$ or
 \begin{equation}
 \label{gyrosym}
 \partial_\gamma \omega_c=0
 \end{equation}
  also for  general magnetic fields. This  assumption is relevant because,  from (\ref{Psi0}), $\Psi=\Psi(t,X,\rho_L)=-(m/e)\omega_c \rho_L^2$. Moreover, from the representation (\ref{clebsh}), $\Psi$ can be identified with the toroidal magnetic flux, and the \emph{Larmor} radius $\rho_L$, can be identified with the radial coordinate used to label the magnetic flux surfaces, $\rho_L=r$ (see the introduction of section~\ref{sec:sec6}). If $X$ lies on the magnetic axis, it  seems reasonable identifying the gyro-angle $\gamma$ to be $\gamma=\vartheta-\phi/q$, as if considered a generalized poloidal angle for the magnetic field expressed in toroidal flux coordinates.

\section{Magnetic moment and drift velocity }
\label{sec:sec8}

In electrodynamics the magnetic moment $\mu$ is considered an adiabatic invariant of motion. Such crucial concept is in conflict with the present purpose of obtaining a non perturbative description of motion. Indeed, the adjective ''adiabatic" refers to an intrinsically perturbative procedure, again. The magnetic moment for a gyro-particle has been already considered in (\ref{gyrosol2}), explicitly rewritten below
\begin{equation}
\mu=-b\cdot\rho \times [\sigma+(e/m)A] = -(e/m) \rho_L^2 \sigma \cdot \nabla \gamma -(e/m)^2\Psi,
\end{equation}
as the conserved momentum associated to the gyro-symmetry, which allows the orbit of a gyro-particle  to close. For the charged particle expressed by (\ref{chargeCoord}), $\mu$ can be similarly defined as proportional to the conjugate momentum (in \emph{hamiltonian} sense) of the ignorable coordinate $\gamma$. \\  
The \emph{cyclotron} frequency  has already been defined as $\omega_c=\dot \gamma$, with $\partial_\gamma \omega_c=0$.\\
 The \emph{Hamiltonian}, $\mathcal{H}$, associated to the description of the particle motion  is
\begin{eqnarray}
\mathcal{H}(t,x,p)&=&\frac{[p-(e/m)(A+\delta_\rho A)]^2}{2}+\frac{e}{m}(\Phi+\delta_\rho \Phi)=\\
&=&\frac{[P-(e/m)A]^2}{2}+\frac{\sigma^2}{2}+\sigma \cdot \left(P-\frac{eA}{m} \right)+\frac{e}{m}(\Phi+\delta_\rho \Phi)
\end{eqnarray}
It is now possible to define the magnetic moment as
\begin{equation}
\label{magmo}
\mu \equiv \frac{e\sigma^2}{2m\omega_c}+\frac{e\sigma \cdot V}{m\omega_c}+\left(\frac{e}{m}\right)^2\frac{\delta_\rho \Phi}{\omega_c},
\end{equation}
so that the \emph{Hamiltonian} becomes
\begin{equation}
\mathcal{H}(t,x,p)=H(t,X,P,(m/e)\gamma,\mu)=\frac{[P-(e/m)A]^2}{2}+\frac{e}{m}\Phi+\frac{m}{e}\mu \omega_c.
\end{equation}
and the motion is described by the canonical coordinates $(X,P)$ and $((m/e)\gamma,\mu)$. The \emph{Hamiltonian} equations for the last couple of coordinates are trivial:
\begin{equation}
\begin{cases}
\label{ham1}
&(m/e)\dot \gamma = \partial_\mu H=(m/e)\omega_c, \\
&\dot \mu= -(m/e)\partial_\gamma H=0,
\end{cases}
\end{equation}
with the desired conservation rule.
Moreover, for the coordinates $(X,P)$:
\begin{equation}
\label{hameq}
\begin{cases}
&\dot X= \nabla_P H=P-(e/m)A, \\
&\dot P = -\nabla H= (e/m)[P-(e/m)A]\cdot \nabla A+(e/m)[P-(e/m)A]\times \nabla \times A+\\
&\qquad \qquad \qquad - (e/m) \nabla \Phi-(m/e)\mu\nabla \omega_c.
\end{cases}
\end{equation}
The last equation, after some algebra, is nothing else that the \emph{Lorentz} force law with an effective electric field given by $E_{\mathrm{eff}}=E-(m/e) \mu \nabla \omega_c$:\begin{equation}
\dot V=\frac{e}{m}(E_{\mathrm{eff}} + V \times B).
\end{equation}
There is not any canonical neither non canonical transformation from $(x,p)\rightarrow (X,P,\gamma,\mu)$ because the dimension of the two spaces is different. Fortunately, this can be done without problems advancing $P$ to be a field $P=P(t,X)$. In this way, also $V$ becomes a field $V=V(t,X)$ and the first \emph{Hamilton}'s equation in (\ref{hameq}) gives the starting equation (\ref{first}) of the present work.

 When $V\sim eB/m$, the drift velocity, $V_D=\lambda \nabla \times V$, if explicitly computed, shows we are in the right direction. Starting from the \emph{curvature} vector $\kappa_b$ of $V$, being $\kappa_b=b \cdot \nabla b=-b\times \nabla \times b$, then 
\begin{equation}
\nabla \times b=b \times \kappa_b+ b  b \cdot \nabla \times b.
\end{equation}
The drift velocity becomes:
\begin{equation}
V_D=\lambda \left\{\nabla |V|_s \times b +|V|_s \left( b \times \kappa_b+ b  b \cdot \nabla \times b\right) \right\}.
\end{equation} 
which can be rewritten, for $V/\lambda=\omega_t b$, or $|V|_s=\lambda \omega_t$, and $V^2/2=\varepsilon -(e/m)\Phi-(m/e)\mu \omega_c$,  as:
\begin{equation}
V_D= \frac{\nabla \varepsilon \times b}{\omega_t}  + \frac{1}{\omega_t}b\times\left[\nabla \Phi + \mu \nabla (m\omega_c/e) +V^2 \kappa_b\right]+\frac{V}{\omega_t} V\cdot \nabla \times b.
\end{equation}
Substituting $\nabla \varepsilon=-\partial_t p=-\partial_t [p -(e/m) A]-(e/m)\partial_t A$, $V_D$ becomes:
 \begin{equation}
 \label{DriftV}
V_D= \frac{E \times b}{(m/e)\omega_t} +\frac{\mu}{\omega_t} b \times \nabla (m\omega_c/e) +\frac{V^2}{\omega_t} b \times \kappa_b +\frac{1}{\omega_t}b\times \partial_t [p-(e/m)A]+\frac{V}{\omega_t} V\cdot \nabla \times b.
\end{equation}
 The drift velocity $V_D$ is the sum of five terms, respectively called: the $E \times B$-like drift, the $\mu \nabla |B|$-like drift, the curvature-like drift, the $\partial_t |B|$-like drift and the Ba\~nos term. \\
The velocity $V_D$ is identical in form to the drift velocity computed at the lowest order in standard gyrokinetics \cite{Hazeltine}. Respect to the standard gyro-center drift velocity, $e|B|/m$ must be replaced by $\omega_c$ and/or by $\omega_t$, whilst $b^{(0)}$, the unit vector in the $B$ direction, must be substituted with $b$, parallel to $V$. This means that all  results obtained up to now  with a drift velocity computed at lowest order,  can be easily extended and considered almost right at all orders if you give a different interpretation to the ambient magnetic field, as if you are considering another scenario, surely not so different, but another one. Such remark refers mainly to the laboratory plasma modeling, where there is an optimal control on the equilibrium magnetic field; \emph{e.g.} in fusion devices, where, among other, the configuration of the ambient magnetic field is fundamental for the stability properties of plasma and its  confinement.  

It is worth noticing that it is possible to obtain the $\partial_t |B|$-like drift considering $\nabla \varepsilon =0$ and adding to the drift velocity the term $b \times \partial_t p/\omega_t$. This procedure is the same outlined in section~\ref{sec:sec2} when it is considered an electric field given by $E=E_\mathrm{eff}+E_t$ and, for the present case, $E_t=(e/m)\partial_t p$. Setting $\nabla \varepsilon =0$ means considering $X$ and $\varepsilon$ as independent coordinates. There is an advantage on describing the system through the coordinates $(t,X,\varepsilon,\mu,\gamma)$. Indeed, the charge in a general magnetic field can be described either in canonical coordinates $(x,p)$ as in non canonical coordinates $(X,\varepsilon,\mu,\gamma)$. However, those coordinates can be, sometime, recast into canonical one as in the toroidal symmetry case. In such case, it is possible write $X$ as function of $X=X(\psi,\mathcal{P}_\phi,(e/m)\phi)$, as done in \cite{me} and the motion can be described by the canonical pairs of variables: $(t,\varepsilon)$, $((e/m)\phi,\mathcal{P}_\phi)$ and $((m/e)\gamma,\mu)$;  considering $\psi$ as a time-like coordinate, as precedently suggested, and the conjugate to $\psi$ momentum, $p_\psi=p_\psi(t,(e/m)\phi,(m/e)\gamma,\varepsilon,\mathcal{P}_\phi,\mu)$, as \emph{Hamilton}'s function. 
 Such canonical description is postponed to a future work. However, the same set of coordinates, $(t,X,\varepsilon,\mu,\gamma)$, is used for obtaining the gyrokinetic equation, in the next section.
 To proceed and  obtain the gyrokinetic equation it could be helpful to explicitly express the magnetic moment and the drift velocity for a constant magnetic field, in GC coordinates:
 \begin{equation}
\label{magmoGC}
\mu_0 = \frac{e\sigma^2}{2m\omega_{c0}}+\frac{e\sigma \cdot U}{m\omega_{c0}}+\left(\frac{e}{m}\right)^2\frac{\delta_\rho \Phi_0}{\omega_{c0}},
\end{equation}
and
 \begin{equation}
 \label{DriftU}
U_D= \frac{b \times \nabla \Phi_0}{(m/e)\omega_{t0}} +\frac{\mu_0}{\omega_{t0}} b \times \nabla (m\omega_{c0}/e) +\frac{U^2}{\omega_{t0}} b \times \kappa_b +\frac{U}{\omega_{t0}} U\cdot \nabla \times b,
\end{equation}
respectively.
 
\section{Gyrokinetic equation}
\label{sec:sec9}

The gyrokinetic equation is chosen as the first use of the present approach. The gyrokinetic equation is a fundamental equation for describing plasma behavior  in a magnetic field. Up to now, it has been studied the single particle motion in general magnetic fields, whilst the gyrokinetic equation analyzes a macroscopic system and it is inherent to the many-body physics. Thanks to the \emph{Boltzmann} equation you can distinguish what depends on the dynamic of the single charge from the rest, The distribution function and the \emph{collision operator}, $C$. 
The collision operator applied to the distribution function $f=f(t,x,v)$ can be chosen to be the \emph{modified Landau} collision operator:
\begin{equation}
 \label{LandauC}
C_L=\frac{\gamma_C}{2} \nabla_v \cdot \int d^3 v^\prime \, \frac{1}{|u_p|}\left( \mathbb{1} - \frac{u_p u_p}{u_p^2} \right) \cdot \left( f(t,x,v^\prime) \nabla_v f(t,x,v) - f(t,x,v) \nabla_{v^\prime} f(t,x,v^\prime) \right).
\end{equation}
being $\gamma_C$ a constant which is known for the \emph{Coulomb} collision case, and $u_p$ given by \cite{me3}:
\begin{equation}
\label{uplasma}
u_{p}=v-v^\prime +\frac{eT_w}{m\Delta^2_\mu} \left[ \frac{\sigma}{\omega_c}\frac{2\left( \mu^2 -s_0^2\right)-\Delta_\mu^2}{\mu+s_0}-\frac{\sigma^\prime}{\omega^\prime_c}\frac{2\left( \mu^{\prime 2} -s_0^2\right)-\Delta_\mu^2}{\mu^\prime+s_0}\right],
\end{equation} 
 to allow the system relaxing towards the following equilibrium distribution function \cite{me3}:
 \begin{equation}
 \label{feq}
f_{eq}(\mathcal{P}_\phi,\varepsilon,\mu)=N\frac{1+\mu/s_0}{\pi \Delta_\mu T_w \Delta_{P_\phi}}\exp \left[ -\frac{(\mathcal{P}_\phi-\mathcal{P}_{\phi 0})^2}{\Delta_{P_\phi}^2} \right] \exp\left[ -\frac{\varepsilon}{T_w}\right]\exp  \left[ - \frac{(\mu-s_0)^2}{\Delta_\mu^2}\right],
\end{equation}
being $N, s_0, \Delta_\mu, T_w, \mathcal{P}_{\phi 0}$ and $\Delta_{P_\phi}$ constants parameters.

However, hereinafter, $C$ and $f_{eq}$ are left undetermined. This means that the gyrokinetic equation we get to show cannot be exactly solved, here. The main difference with other derivations is that the gyrokinetic equation is here obtained non-perturbatively, which means that the single particle motion is described without approximation, using the exact magnetic field. Perhaps inevitably, the solution of the gyrokinetic equation have to be expressed perturbatively.\\

The form of gyrokinetic equation depends on the set of coordinates used to represent the charge  motion. The  gyrocenter coordinates $t, X, \varepsilon, \mu$ and $\gamma$ are the present choice. The time derivative of those variables are already known, from (\ref{first}), (\ref{dote}) and (\ref{ham1}):
\begin{equation}
\label{timeder}
\dot X=V, \quad \dot{\varepsilon}=\frac{e}{m}\left[\partial_t \Phi(t,X+\rho) -(V+\sigma)\cdot \partial_t A(t,X+\rho)\right],\quad \dot{\mu}=0,\quad \dot \gamma= \omega_c.
\end{equation}
The \emph{Boltzmann} equation for a distribution function $f$ is 
\begin{equation}
\label{Boltz}
\dot f=C(f),
\end{equation}
or, more explicitly, 
\begin{equation}
\label{Boltz}
\partial_t f+V\cdot \nabla f+\frac{e}{m}\left[\partial_t \Phi(t,X+\rho) -(V+\sigma)\cdot \partial_t A(t,X+\rho)\right]\partial_\varepsilon f+\omega_c \partial_\gamma f=C(f).
\end{equation}
The initial distribution function is assumed to satisfy \emph{Boltzmann} equation: $\dot f_0 = C(f_0)$. For stability studies, it is often considered  an equilibrium distribution function, like $f_{eq}$ in (\ref{feq}), which is a particular solution of the \emph{Boltzamann} equation, being $\dot f_{eq} =C(f_{eq})=0$, separately. The evolution of plasma is obtained by the behavior of the distribution function and the behavior of  fields. The \emph{Boltzmann} equation gives rise to an initial value problem: it is studied the system evolution from an  initial state, the one defined at time $t=0$. All the increments, of the distribution function and fields, are referred to the initial values as below explicitly written: 
 \begin{eqnarray}
 f(t,x,v)&=&f_0(0,Q,\varepsilon_0,\mu_0) + \delta f(t,X,\varepsilon,\mu,\gamma), \\
 \Phi(t,x)&=&\Phi_0(0,x)+\delta \Phi(t,x), \\
 A(t,x)&=&A_0(0,x)+\delta A(t,x),
 \end{eqnarray} 
 being $Q$ the GC position, and $\varepsilon_0=U^2/2+e\Phi_0/m+(m/e)\mu_0 \omega_{c0}$, the energy at $t=0$, with $U$ the GC velocity, $\mu_0$ and $\omega_{c0}$, the magnetic moment and the cyclotron frequency, respectively, always at $t=0$. 
 It is here assumed that the initial distribution function satisfies the \emph{Boltzmann} equation: $\dot f_0 = C(f_0)$ and that $\partial_\gamma f_0=0$. Due to the latter condition,  $f_0$ is said to be the initial distribution function of GCs (instead of particles). The \emph{Boltzmann} equation for $f_0$ is particularly simple:
 \begin{equation}
 \label{BoltzGC}
 U\cdot \nabla f_0 = C(f_0),
 \end{equation}
 being $\partial_t f_0=\dot{\varepsilon}_0=\dot{\mu_0}=0$.\\The gyrokinetic equation describes the functional form of $\delta f$ in terms of initial fields and the initial distribution function, supposed given.
Using (\ref{BoltzGC}), (\ref{Boltz}) and (\ref{BoltzGC}), the \emph{Boltzmann} equation is rewritten as
 \begin{eqnarray}
\label{BoltzdeltaF}
\nonumber
&&\partial_t \delta f+V \cdot \nabla \delta f+\omega_c\partial_\gamma \delta f+\delta_\xi U \cdot \nabla f_0+\frac{e}{m}(\partial_t \delta \Phi -V\cdot \partial_t \delta A -\sigma \cdot \partial_t \delta A  )\partial_\varepsilon f_0=\\
&&=\delta_{\delta f} C-\frac{e}{m}(\partial_t \delta \Phi -V\cdot \partial_t \delta A -\sigma \cdot \partial_t \delta A  )\partial_\varepsilon \delta f,
\end{eqnarray}
being $\delta_{\delta f} C= C(f)-C(f_0)$. In this way it has been moved on the RHS the non-linear terms leaving on the LHS all the linear terms. The RHS can be computed only if $C$ is substituted with an analytical operator, as can be $C_L$ in (\ref{LandauC}), but here $C$ is leaved undetermined and the RHS is indicated as ''$-\nu_{NL}\delta f$", being $\nu_{NL}$ a symbol. Explicitly:
\begin{equation}
\label{BoltzdeltaF}
\partial_t \delta f+V \cdot \nabla \delta f+\omega_c\partial_\gamma \delta f+\delta_\xi U \cdot \nabla f_0+\frac{e}{m}(\partial_t \delta \Phi -V\cdot \partial_t \delta A -\sigma \cdot \partial_t \delta A  )\partial_\varepsilon f_0=-\nu_{NL} \delta f.
\end{equation} 
The equation (\ref{BoltzdeltaF})  is the \emph{Boltzamann} equation for $\delta f$ which represents all the charges in the distribution function that, at time $t\neq0$, cannot be anymore represented by $f_0$. Such equation is studied in the dual \emph{Fourier} space. The \emph{Fourier} transform in $(t,X,\gamma)$ of $\delta f$ is $\delta \tilde f_l$ defined as:
\begin{equation}
\delta f = \sum_{l=-\infty}^{\infty} \,\int \, d^3k d\omega e^{i\omega t}e^{-ik \cdot X}e^{-i l\gamma}\delta \tilde f_l (\omega,k),
\end{equation}
with the function $\tilde f_l (\omega,k)$ depending also on energy, magnetic momentum and GC displacement vector, even if not explicitly indicated.
The \emph{Fourier} transform of e.m. potentials are:
\begin{equation}
\delta \Phi(t,X+\rho)=\int \, d^3k d\omega e^{i\omega t}e^{-ik\cdot (X+\rho)} \delta \tilde \Phi(\omega,k),
\end{equation}
\begin{equation}
\delta A(t,X+\rho)= \int \, d^3k d\omega e^{i\omega t}e^{-ik \cdot (X+\rho)}\delta \tilde A (\omega,k).
\end{equation}
At least, the \emph{Fourier} transform of $\delta_\xi U$ is
\begin{equation}
\delta_\xi U= \int \, d^3k d\omega e^{i\omega t}e^{-ik \cdot Q}\delta_\xi \tilde U (\omega,k).
\end{equation}
The equation (\ref{BoltzdeltaF}),  in the \emph{Fourier} dual space, is
 \begin{eqnarray}
\label{BoltzFourier}
\nonumber
&&\partial_t \delta f+V \cdot \nabla \delta f+\omega_c\partial_\gamma \delta f+\nu_{NL} \delta f+\\
\nonumber
&&+\delta_\xi U \cdot \nabla f_0+\frac{e}{m}(\partial_t \delta \Phi -V\cdot \partial_t \delta A -\sigma \cdot \partial_t \delta A  )\partial_\varepsilon f_0=\\
\nonumber
&&= i \sum_{l=-\infty}^{\infty} \,\int \, d^3k d\omega e^{i\omega t}e^{-ik \cdot X}e^{-i l\gamma}(\omega-k\cdot V-l \omega_c-i\nu_{NL})\delta \tilde f_l+\\
\nonumber
&&+\,\int \, d^3k d\omega e^{i\omega t}e^{-ik \cdot Q} \delta_\xi \tilde U \cdot \nabla f_0 +\\
&&+i \,\int \, d^3k d\omega e^{i\omega t}e^{-ik \cdot X}(e/m)\omega  e^{-ik \cdot \rho}( \delta \tilde \Phi -V\cdot \delta \tilde A- \sigma \cdot \delta \tilde A)\partial_\varepsilon f_0=0.
\end{eqnarray}
The same equation is below written with the substitution $\sigma=\omega_c \rho \times b$, being $\partial_\gamma \omega_c=0$, from (\ref{gyrosym}), and the tern of orthogonal unit vectors $e_\rho \cdot b \times e_\gamma=1$ (computed at time $t$ and position $x=X+\rho$).
When equation (\ref{BoltzFourier}) is anti transformed in $\gamma$, then 
\begin{eqnarray}
\label{BoltzFourier3}
\nonumber
&&\oint \frac{d\gamma}{2\pi} e^{il^\prime \gamma} [\partial_t \delta f+V \cdot \nabla \delta f+\omega_c\partial_\gamma \delta f+\frac{e}{m}(\partial_t \delta \Phi -V\cdot \partial_t \delta A -\sigma \cdot \partial_t \delta A  )\partial_\varepsilon f_0-\nu_{NL}\delta f]=\\
\nonumber
&&= i \,\int \, d^3k d\omega e^{i\omega t}e^{-ik \cdot X} (\omega-k\cdot V-l^\prime \omega_c-i\nu_{NL})\delta \tilde f_{l^\prime}+\delta_{l^\prime \,0} \int \, d^3k d\omega e^{i\omega t}e^{-ik \cdot Q}  \delta_\xi \tilde U \cdot \nabla f_0 +\\
&&+ i \,\int \, d^3k d\omega e^{i\omega t}e^{-ik \cdot X} \frac{e\omega}{m} [ (\delta \tilde \Phi -V\cdot \delta \tilde A)- \omega_c b \times \delta \tilde A \cdot   \nabla_k] \partial_\varepsilon f_0 \oint \, \frac{d\gamma}{2\pi} e^{i l^\prime \gamma-ik \cdot \rho}=0.
\end{eqnarray}
The \emph{wave number} $k$ varies in the dual space but it can be conveniently chosen to be expressed in cylindrical coordinates: $k=k_b b+k_t=k_b b +|k_t| [e_\rho \sin (\gamma-\gamma_k )+e_\gamma \cos (\gamma-\gamma_k)]$, being $b$ the \emph{longitudinal} (parallel to $\Omega$) unit vector, $k_b$ the \emph{longitudinal} wave number component, $|k_t|$ the amplitude of the \emph{transversal} wave number component and $\gamma_k$ an arbitrary gyro-angle.
The \emph{cylindrical Bessel} functions, $J_l$, are  defined by
\begin{equation}
\label{bessel}
J_l(|k_t|\rho_L)\equiv \frac{1}{2\pi } \oint \, d\gamma e^{i l (\gamma-\gamma_k)-i|k_t| \rho_L \sin(\gamma-\gamma_k)}= \frac{ e^{-il\gamma_k}}{2\pi }\oint \, d\gamma e^{i l \gamma-ik \cdot \rho},
\end{equation}
whose partial derivatives respect to $k$ are
\begin{equation}
\nabla_k J_l(|k_t|\rho_L)=\nabla_k |k_t| \cdot \partial_{|k_t|} J_l(|k_t|\rho_L)=\frac{\rho_L k_t}{|k_t|}J^\prime_l(|k_t|\rho_L),
\end{equation}
and satisfying the following property:
\begin{equation}
J^\prime_l=\begin{cases} (J_{l-1}-J_{l+1})/2, & \mbox{  if } l \neq 0 \\-J_1, & \mbox{ if } l = 0.
\end{cases}
\end{equation}
Equation (\ref{BoltzFourier3}), with the use of the cylindrical \emph{Bessel} functions, is rewritten as  
\begin{eqnarray}
\label{BoltzFourier4}
\nonumber
&& i \,\int \, d^3k d\omega e^{i\omega t}e^{-ik \cdot X} (\omega-k\cdot V-l^\prime \omega_c-i\nu_{NL})\delta \tilde f_{l^\prime}= \delta_{l^\prime \,0} \int \, d^3k d\omega e^{i\omega t}e^{-ik \cdot Q}  \delta_\xi \tilde U \cdot \nabla f_0+\\
&&- i\,\int \, d^3k d\omega e^{i\omega t}e^{-ik \cdot X}e^{il^\prime \gamma_k}\frac{e\omega}{m} [(\delta \tilde \Phi -V\cdot \delta \tilde A) J_{l^\prime}- \frac{\rho_L\omega_c}{|k_t|} b \times \delta \tilde A \cdot  k J^\prime_{l^\prime}] \partial_\varepsilon f_0.
\end{eqnarray}
Equation (\ref{BoltzFourier4}) is always the \emph{Boltzmann} equation applied to $\delta f$ and described in the Fourier dual space. The gyrokinetic equation is the same equation with $l^\prime=0$:
 \begin{eqnarray}
\label{gyrokin}
\nonumber
&& i \,\int \, d^3k d\omega e^{i\omega t}e^{-ik \cdot X} (\omega-k\cdot V-i\nu_{NL})\delta \tilde f_0=\\
&&=\partial_t \delta f_0+V \cdot \nabla \delta f_0 +\nu_{NL} \delta f_0= \int \, d^3k d\omega e^{i\omega t}e^{-ik \cdot Q}  \delta_\xi \tilde U \cdot \nabla f_0+\\
\nonumber
&& -i\,\int \, d^3k d\omega e^{i\omega t}e^{-ik \cdot X}\frac{e\omega}{m} [(\delta \tilde \Phi -V\cdot \delta \tilde A) J_0+ \frac{\rho_L\omega_c}{|k_t|} b \times \delta \tilde A \cdot  k J_1] \partial_\varepsilon f_0,
\end{eqnarray}
where
\begin{equation}
 \delta f_0=\oint \frac{d\gamma}{2\pi} \delta f,
 \end{equation}
 is the gyro-average of $\delta f$ and it represents the gyro-centers in $\delta f$.
The solution for $\delta \tilde{f}_{l^\prime}$ is
\begin{eqnarray}
\nonumber
&&\delta \tilde f_{l^\prime}= -\frac{e\omega e^{il^\prime \gamma_k}}{m}\frac{[(\delta \tilde \Phi -V\cdot \delta \tilde A) J_{l^\prime}- (\rho_L\omega_c/|k_t|) k \times b \cdot \delta \tilde A  J^\prime_{l^\prime}] \partial_\varepsilon f_0}{\omega-k\cdot V-l^\prime \omega_c-i\nu_{NL}}+\\
&&\qquad - e^{il^\prime \gamma_k}\frac{i\delta_{l^\prime\,0} e^{ik\cdot \xi}\delta_\xi \tilde U\cdot \nabla f_0}{\omega-k\cdot V-l^\prime \omega_c-i\nu_{NL}}
\end{eqnarray}
In particular, for $l^\prime=0$, the solution $\delta \tilde f_0$ of the gyrokinetic equation is
\begin{eqnarray}
\label{gyrokinsol}
\nonumber
&&\delta \tilde f_0=- (e\omega/m)\frac{[(\delta \tilde \Phi -V\cdot \delta \tilde A) J_{0}+ (\rho_L\omega_c/|k_t|) k \times b \cdot \delta \tilde A  J_1] \partial_\varepsilon f_0}{\omega-k\cdot V-i\nu_{NL}}+\\
&& \qquad - i \frac{e^{ik\cdot \xi}\delta_\xi U \cdot \nabla f_0}{\omega-k\cdot V-i\nu_{NL}}.
\end{eqnarray}
In real space, the solution of the \emph{Boltzmann} equation is
\begin{eqnarray}
\label{Boltzsol}
&&\delta f  =-\sum_{l=-\infty}^{\infty} \,\int \, d^3k d\omega  e^{i\omega t}e^{-ik \cdot X} e^{-i l(\gamma-\gamma_k)} \frac{e\omega}{m}\cdot\\
\nonumber
&&\cdot  \frac{[(\delta \tilde \Phi -V\cdot \delta \tilde A) J_l- (\rho_L\omega_c/|k_t|) k \times b \cdot \delta \tilde A  J^\prime_l] \partial_\varepsilon f_0+i(m/e)\delta_{l\,0} e^{ik\cdot \xi}(\delta_\xi U/\omega)\cdot \nabla f_0}{\omega-k\cdot V-l \omega_c-i\nu_{NL}}.
\end{eqnarray}
Such solution is formal in the sense that it employes the \emph{symbol} $\nu_{NL}$ which is not a well defined mathematical object. If $C$ is known, then the problem could be attacked with the (perturbative) functional analysis and, maybe, solved. There is another limiting behavior of the system in which the solution can be given without explicitly knowing the collisional operator. This happens at low collisionality (if $C=0$ the equation (\ref{Boltz}) is said \emph{collisionless Boltzmann equation}) when the magnetic field is slowly varying and the initial distribution function $f_0$ is an equilibrium one: $f_0 \to f_{eq}$. In such case the system is said to be at \emph{marginal stability}. In this case, the distribution function $\delta f$ is very small if compared to $f_{eq}$; moreover, $V \to U$, or $\xi \to 0$, and $\nu_{NL}\to \nu$ becomes a real number. At marginal stability, the solution (\ref{gyrokinsol}) for the gyrokinetic equation (\ref{gyrokin}) is well defined (if $\nu$ replaces $\nu_{NL}$ and $\xi=0$):
\begin{equation}
\label{gyrokinsolMarginal}
\delta \tilde f_0 \mid_{m.s.}=- (e\omega/m)\frac{[(\delta \tilde \Phi -U\cdot \delta \tilde A) J_{0}+ (\rho_L\omega_c/|k_t|) k \times b \cdot \delta \tilde A  J_1] \partial_\varepsilon f_{eq}}{\omega-k\cdot U-i\nu}\end{equation}
 The same is true for (\ref{Boltzsol}) which is  solution of the Boltzmann equation at \emph{marginal stability}:
 \begin{equation}
\delta \tilde f_{l^\prime} \mid_{m.s.}= -\frac{e\omega e^{il^\prime \gamma_k}}{m}\frac{[(\delta \tilde \Phi -U\cdot \delta \tilde A) J_{l^\prime}- (\rho_L\omega_c/|k_t|) k \times b \cdot \delta \tilde A  J^\prime_{l^\prime}] \partial_\varepsilon f_{eq}}{\omega-k\cdot U-l^\prime \omega_c-i\nu}
\end{equation}
\\
In the general case, the differences between (\ref{gyrokinsol}), where the quantities are not obtained perturbatively, and the standard gyrokinetic equation, \emph{e.g.} in \cite{FriChen}, are mildly overcome if $\partial_\varepsilon f_0$ is written explicitly:
\begin{equation}
\partial_\varepsilon f_0=\partial_\varepsilon \varepsilon_0 \cdot \partial_{\varepsilon_0} f_0+\partial_\varepsilon Q \cdot \nabla f_0+\partial_\varepsilon \mu_0 \cdot \partial_{\mu_0} f_0,
\end{equation}
and, after that, only the first order quantities are considered: $b \to b^{(0)}$, $U \to v_\|$ and $\omega_c \to (e|B|/m)$. However, a one to one correspondence seems to be lost and it should be further investigated. Indeed,  such differences can be very important for stability studies of plasmas, as recently reviewed in \cite{chenzonca14}.

\section{Conclusions}
\label{sec:concl} 

 In the present work it has been shown the possibility of analytically describing the non relativistic motion of  charged particles in general magnetic fields.  The method used to address this problem is the same approach used by the gyro-center transformation but adopting a non perturbative magnetic field description to find new solutions of motion in GC or gyro-center coordinates. The equation (\ref{claudio}) has been easily obtained from combining  the main equation for a dynamic system, (\ref{first}), the \emph{Lorentz} force law and the condition (\ref{conditions}). From the equation (\ref{claudio}) two solutions have been proposed: the GP and the gyro-particle solutions. The first has been characterized to be minimally coupled to the magnetic field, the second has been defined to be maximally coupled to the magnetic field and also to move on a closed orbit. A generic charged particle has been shown to have a velocity expressed as the sum of such particular solutions, (\ref{chargeCoord}) and (\ref{chargeCoord2}), and characterized by the gyro-centers coordinates $(X,V)$, a \emph{Larmor} radius $\rho_L$ and a gyro-angle $\gamma$. The GC description of motion is similar to the gyro-center description but with the gyro-center coordinates $X,V$ computed at a particular time, here settled at $t=0$. The explicit motion description  has been shown in section~\ref{sec:sec6} where the magnetic field has been chosen to have an axial  symmetry. The  axisymmetric toroidal magnetic field, often encountered for describing laboratory plasmas, has been extensively studied obtaining new relevant representations for the velocity of particles and for the toroidal canonical momentum. The symmetry in the gyro-angle has revealed the importance on having defined the gyro-particle as moving on closed orbit. Such condition ensures the existence of a particular \emph{gauge} function (\ref{ggauge}), which is proportional to the product of the magnetic moment of the gyro-particle times the gyro-angle.  In such \emph{gauge}, with the defined coordinates, it has been possible to represent the magnetic field seen by the particle as an axisymmetric magnetic field. The gyro-angle $\gamma$ becomes an ignorable coordinate and the \emph{Larmor} radius is  shown to be proportional to the magnetic flux linked to the closed orbit. It is furnished the definition (\ref{magmo}) of the magnetic moment in gyro-center as in GC coordinates. The magnetic moment is considered an exact constant of motion instead of an adiabatic one, and also the drift velocity is expressed in an exact form (\ref{DriftV}). Such drift velocity representation is particularly simple because it is very similar to the drift velocity developed at the first order in the standard gyro-center transformation. Indeed, it is shown to be constituted by the $E\times B$-like drift, the $\nabla |B|$-like drift, the curvature-like drift, the $\partial_t |B|$-like drift and the \emph{Ba\~nos}-like drift. Once a particular set of GC coordinates $(t,X,\varepsilon, \mu, \gamma)$, is chosen to describe the motion of a charge in a general magnetic field, the first use of the present approach is on deriving the gyrokinetic equation. Such derivation of gyrokinetic equation from the \emph{Boltzmann} equation without explicitly considering the collision operator, is shown without any approximations (neither ordering considerations). The final form (\ref{gyrokin}), which is particularly simple, is formally solved,  (\ref{gyrokinsol}), together with the formal solution, (\ref{Boltzsol}), of the \emph{Boltzmann} equation for the perturbed distribution function $\delta f$. The obtained solution can be considered the correct solution for the \emph{Boltzmann} equation at \emph{marginal stability}. The next steps, concerning the use of the present non perturbative approach to gyrokinetics, should be to address the \emph{Vlasov} equation, where the fluctuation of fields is coupled to $\delta f$ via \emph{Maxwell} equations, and to study its phenomenologies, \emph{e.g.} obtaining the plasma dispersion relation.  

\section*{Acknowledgments}

This work was performed in the frame of the former ENEA-Euratom Contract of Association.


\begin{thebibliography}{99}
\bibitem{footnote1}
It is worth noticing that (\ref{simHam}) is similar to the  \emph{Hamilton}'s equation: $\dot{P}+\nabla \mathcal{E}=0$; 
 moreover, it can be also associated to the energy tensor conservation law.
\bibitem{footnote2}
 The difference respect to the current gyrokinetics framework is that the present approach is non perturbative and, in principle, it is possible to find exact solutions in addition to the perturbative ones.  Exact solutions are better investigated starting from specific expression of $B$, as it will be shown in section~5 for axisymmetric systems.
\bibitem{footnote3}
 From such definition it is possible either that it doesn't exist such reference frame or that, if it exists, it is not unique.
\bibitem{footnoten}
 For such solution, the transverse electric field is zero, being $E_t(t,X+\rho)=-(m/e)(\partial_t p +\nabla \varepsilon)= 0$. Another kind of solution can be obtained from the following system of equations:
\begin{equation}
\nonumber
\begin{cases} \partial_t p+\nabla \varepsilon=kV \times \sigma, \\
\nabla \times V+eB/m=V/\lambda,\\
 \nabla \times \sigma+V/\lambda+ e\delta_\rho B/m=k \sigma,
\end{cases}
\end{equation}
which has the peculiarity that the transverse electric field is perpendicular both to $V$ and to $\sigma$.
\end{thebibliography}
\end{document}